\documentclass[submission,copyright,creativecommons]{eptcs}
\providecommand{\event}{TARK 2017}

\usepackage{booktabs}

\usepackage{verbatim}
\usepackage[usenames,dvipsnames,svgnames,table]{xcolor}
\usepackage{graphicx}
\usepackage{url}
\usepackage{cite}
\usepackage{eurosym}
\usepackage{tikz}
\usetikzlibrary{arrows,matrix}

\newcommand{\up}{\vspace{0cm}}

\usepackage{xcolor}

\title{What Drives People's Choices in Turn-Taking Games, \\if not Game-Theoretic Rationality?}
\author{Sujata Ghosh
\institute{Indian Statistical Institute\\ Chennai, India}
\email{sujata@isichennai.res.in}
\and
Aviad Heifetz
\institute{The Open University of Israel\\ Raanana, Israel}
\email{aviadhe@openu.ac.il}
\and
Rineke Verbrugge
\institute{University of Groningen\\ Groningen, The Netherlands}
\email{L.C.Verbrugge@rug.nl}
\and
Harmen de Weerd
\institute{University of Groningen\\ Groningen, The Netherlands}
\email{h.a.de.weerd@rug.nl}
}
\def\titlerunning{What Drives People's Choices in Turn-Taking Games, if not Game-Theoretic Rationality?}
\def\authorrunning{Sujata Ghosh, Aviad Heifetz, Rineke Verbrugge \& Harmen de Weerd}
\def\copyrightholders{S. Ghosh, A. Heifetz, R. Verbrugge \& H. de Weerd}

\begin{document}


\maketitle


\begin{abstract}
In an earlier experiment, participants played a perfect information game against a computer, which was programmed to deviate often from its backward induction strategy right at the beginning of the game. Participants knew that in each game, the computer was nevertheless optimizing against some belief about the participant's future strategy. In the aggregate, it appeared that participants applied forward induction. However, cardinal effects seemed to play a role as well: a number of participants might have been trying to maximize expected utility.

In order to find out how people really reason in such a game, we designed centipede-like turn-taking games with new payoff structures in order to make such cardinal effects less likely. We ran a new experiment with 50 participants, based on marble drop visualizations of these revised payoff structures. After participants played 48 test games, we asked a number of questions to gauge the participants' reasoning about their own and the opponent's strategy at all decision nodes of a sample game. We also checked how the verbalized strategies fit to the actual choices they made at all their decision points in the 48 test games.

Even though in the aggregate, participants in the new experiment still tend to slightly favor the forward induction choice at their first decision node, 
their verbalized strategies most often depend on their own attitudes towards risk and those they assign to the computer opponent, sometimes in addition to considerations about cooperativeness and competitiveness.
\end{abstract}



\section{Introduction}\label{sec:intro}



In game theory, turn-taking games are represented by game trees referred to as extensive-form games. It is well known that the concept of {\em game-theoretic rationality} provides a canonical approach towards solving extensive-form games with perfect information, namely {\em Backward Induction} (BI), which yields a subgame-perfect equilibrium   (which is a unique  in generic turn-taking games with no pay-off ties)~\cite{perea12,perea07}.  However, BI is often criticized for not taking into account that a player may end up in one particular subgame rather than another subgame, e.g. after a deviation of other players from their BI strategy. Thus, the past moves and reasoning of the players are not taken into consideration, only the future. All players commonly believe in everybody's future rationality, no matter how irrational players' past behavior has already proven. In some cases, such complete lack of consideration of previous moves may be suboptimal for players.


An alternative approach, that of {\em Forward Induction} (FI), does take these previous moves of the opponent(s) under consideration and tries to rationalize the opponent's past behavior for assessing his future moves. Thus, when a player is about to play in a subgame that has been reached due to some strategy of the opponent that is not consistent with both common knowledge (belief) of rationality for each of the players {\em and} his past behavior, the player may still rationalize the opponent's past behavior. She may attribute to her opponent a strategy that is optimal against a possible suboptimal strategy of hers, or   a strategy that is optimal against some rational strategy of hers, which is only optimal against a suboptimal strategy of his, and so on. If the player pursues this kind of rationalizing reasoning to the highest extent possible and reacts accordingly, she ends up choosing what is called an Extensive-Form Rationalizable (EFR) strategy~\cite{pearce84,battigalli96,battigalli97}. For perfect information games, to which we restrict our attention in this paper, see Definition 1 in~\cite{hp14}, which we we will be using here. 


Even though EFR strategies may be distinct from BI strategies (e.g., see~\cite{reny92}, Game 1 in Figure 1), in perfect-information  games without relevant payoff ties it has be been shown that the unique EFR outcome   coincides with the unique BI outcome~\cite{battigalli96,battigalli97,hp14} . In case there are relevant payoff ties, however,  EFR outcomes may   form a---possibly strict---subset of the BI outcomes~\cite{chenmicali11,chenmicali13,perea12,hp14}, see game 3 in Figure 1.




In~\cite{ghv14}, we asked the question:
 {\em Are people inclined to use forward induction when they play a game?} Our
pivotal interest was to examine participants' behavior following a deviation from
BI behavior by their opponent right at the beginning of the game. We designed a Marble Drop game experiment where the participants played several rounds of turn-taking games with the computer. Related research has been done on how people reason in dynamic games, also focusing on forward induction~ \cite{bn08,shahriar14,cachon1996,huck2005,chlass2016}. However, their games are not perfect information, in contrast to our Marble Drop games. In these games, we programmed the computer so as to follow, in each repetition of each game, a strategy which is optimal with respect to some strategy of the human participant. We provided a more intuitive framing of the game trees, inspired by Meijering et al.'s Marble Drop~\cite{meijering2010,meijering2011}. In ~\cite{ghv14}, it turned out that in the aggregate, people's first decisions could be explained as EFR behavior. However, it also seemed that in many cases, cardinal effects could have played a role. Consequently, we wanted to design a new Marble Drop experiment which minimizes this cardinal effect, and in the process, improve certain other aspects of our previous experiment, based on our own findings and discussion with colleagues working in empirical research.


To summarize, the main differences between the current Marble Drop experiment and the one reported in~\cite{ghv14} are as follows. The aim is to have better overall understanding of the participants' behavior: \\

\noindent - At the beginning of the experiment, we explained verbally in a more comprehensive way regarding how the computer agent was programmed (including ``Imagine that you are playing against a different computer opponent each time" and ``the computer thinks that you already have a plan for that game, and it plays the best response to the plan it thinks that you have for that game, however, the computer does not learn from previous games and does not take into account your choices during the previous games".)

\noindent - We changed the payoff structures of the experimental games in order to minimize cardinal effects on the behavior of the participants.

\noindent - We modified the questions asked to the Group A and Group B participants to direct some of the participants towards strategic reasoning while playing the games and note the effects of such prompts.

\noindent - We asked more structured and pointed questions about the participants' decision-making process at the end of the experiment to bring out the reasoning behind their behaviors in a more explicit manner.

\noindent -We changed the monetary incentive so that a wealth effect would be eradicated. By basing payment on number of marbles in a randomly drawn game, the participants perceived a clear difference between earning $k$ and $k+1$ marbles in a game (not just 5 cents as in the previous experiment, but 3.75 euros).\\


\noindent We mainly focused on the following questions:\\

\noindent 1) Do participants now more clearly play according to FI?

\noindent 2) If not, what are they actually doing? What roles are played by risk attitudes and cooperativeness versus competitiveness?

\noindent 3) Can they be reasonably divided into types of players?\\


Additionally, we are interested in whether people take the perspective of their opponent and make use of \emph{theory of mind} while playing the Marble Drop experiment. Theory of mind refers to the ability to reason about unobservable mental content of others, such as beliefs, desires, or goals~\cite{Premack1978}. This theory of mind ability can even be used to reason about the theory of mind of others, and reason about the way others reason about beliefs and goals. This \emph{second-order theory of mind} allows us to understand sentences such as ``Alice \emph{knows} that Bob \emph{wants} to get six marbles'', and use this to adjust our predictions of the behavior of Alice.

The rest of this paper is structured as follows. In Section 2, we explain the new experimental games and the differences with the previous experiment~\cite{ghv14}. Section 3 presents the experimental results about the participants' decisions in the games as well as their reports about their own reasoning behind their behaviors. Section 4 discusses the results and provides suggestions for future research.


\section{A Marble Drop Experiment}\label{sec:expt} 


In~\cite{ghv14}, we designed a marble drop game experiment to investigate whether people are inclined to use Forward Induction (FI) when they play dynamic perfect-information games. The participants played 8 rounds of turn-based games against a computer opponent, repeating in each round a set of 6 games, distinct in terms of payoff structures. By letting participants play against a computer opponent, we ensure that each participant encounters the same situations, which allows us to eliminate \emph{variability due to the strategy of the opponent} in our analysis.  In these two-player games, the players played alternately. Let $C$ denote the computer, and let $P$ denote the participant. 
In four of the games the computer plays first, followed by the participant, and each of the players can play at two decision nodes. In the remaining two games, which are truncated versions of two of the games described earlier, the participant gets first chance to move. It appeared that participants did apply FI in playing these games, that is, in all likelihood the participants responded in a way which is optimal with respect to the conjecture that the computer is after a larger prize than the one it has foregone, even when this necessarily meant that the computer has attributed future irrationality to the participant when the computer made the first move in the games. However, a closer look at the individual choices revealed that cardinal effects probably played a role in these choices.

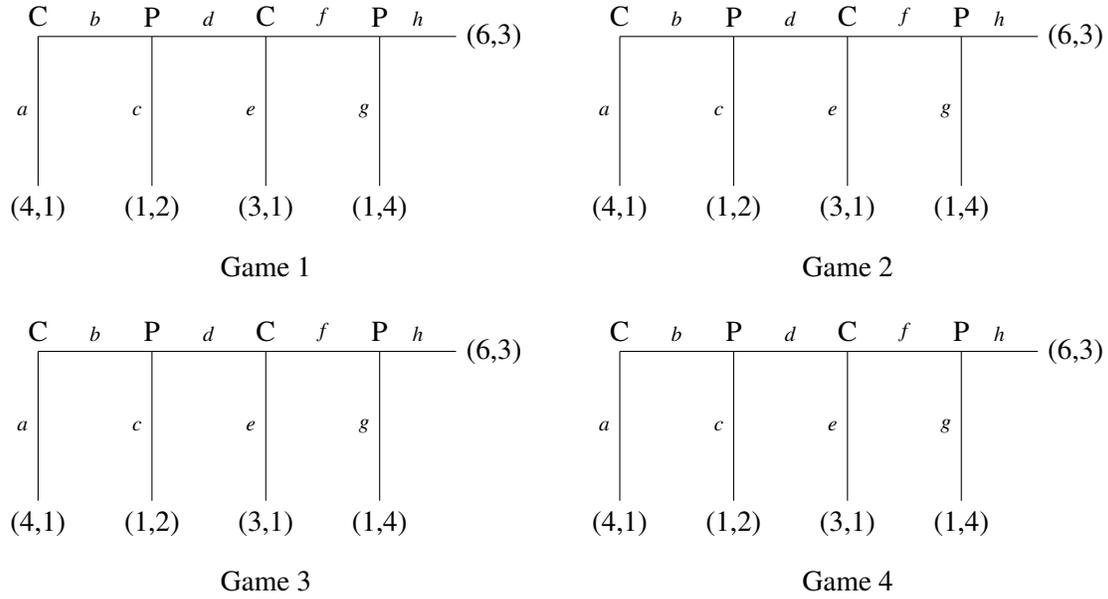
\begin{figure*}[t]
\begin{tabular}{cc}
\begin{tikzpicture}[
dot/.style={shape=circle,fill=black,minimum size=0pt,
	inner sep=0pt,outer sep=0pt},
]
\matrix[matrix of nodes, column sep=3ex, row sep=4ex] {
	|[dot,label=above:C] (p1)| {} & |[dot,label=above:P] (p2)| {} & |[dot,label=above:C] (p3)| {} & |[dot,label=above:P] (p4)| {} & |[dot, label=right:{(6,3)}] (p5)| {}\\[1cm]
	|[dot,label=below:{(4,1)}] (p6)| {}&|[dot,label=below:{(1,2)}] (p7)| {} &|[dot,label=below:{(3,1)}] (p8)| {} & |[dot,label=below:{(1,4)}] (p9)| {} & \\
};
\draw (p1) edge node[above]{\scriptsize{$b$}} (p2);
\draw (p2) edge node[above]{\scriptsize{$d$}} (p3);
\draw (p3) edge node[above]{\scriptsize{$f$}} (p4);
\draw (p4) edge node[above]{\scriptsize{$h$}} (p5);
\draw (p1) edge node[left]{\scriptsize{$a$}} (p6);
\draw (p2) edge node[left]{\scriptsize{$c$}} (p7);
\draw (p3) edge node[left]{\scriptsize{$e$}} (p8);
\draw (p4) edge node[left]{\scriptsize{$g$}} (p9);
\end{tikzpicture}&
\begin{tikzpicture}[
dot/.style={shape=circle,fill=black,minimum size=0pt,
	inner sep=0pt,outer sep=0pt},
]
\matrix[matrix of nodes, column sep=3ex, row sep=4ex] {
	|[dot,label=above:C] (p1)| {} & |[dot,label=above:P] (p2)| {} & |[dot,label=above:C] (p3)| {} & |[dot,label=above:P] (p4)| {} & |[dot, label=right:{(6,3)}] (p5)| {}\\[1cm]
	|[dot,label=below:{(4,1)}] (p6)| {}&|[dot,label=below:{(1,2)}] (p7)| {} &|[dot,label=below:{(3,1)}] (p8)| {} & |[dot,label=below:{(1,4)}] (p9)| {} & \\
};
\draw (p1) edge node[above]{\scriptsize{$b$}} (p2);
\draw (p2) edge node[above]{\scriptsize{$d$}} (p3);
\draw (p3) edge node[above]{\scriptsize{$f$}} (p4);
\draw (p4) edge node[above]{\scriptsize{$h$}} (p5);
\draw (p1) edge node[left]{\scriptsize{$a$}} (p6);
\draw (p2) edge node[left]{\scriptsize{$c$}} (p7);
\draw (p3) edge node[left]{\scriptsize{$e$}} (p8);
\draw (p4) edge node[left]{\scriptsize{$g$}} (p9);
\end{tikzpicture}\\
Game 1&Game 2\\[2mm]
\begin{tikzpicture}[
dot/.style={shape=circle,fill=black,minimum size=0pt,
	inner sep=0pt,outer sep=0pt},
]
\matrix[matrix of nodes, column sep=3ex, row sep=4ex] {
	|[dot,label=above:C] (p1)| {} & |[dot,label=above:P] (p2)| {} & |[dot,label=above:C] (p3)| {} & |[dot,label=above:P] (p4)| {} & |[dot, label=right:{(6,3)}] (p5)| {}\\[1cm]
	|[dot,label=below:{(4,1)}] (p6)| {}&|[dot,label=below:{(1,2)}] (p7)| {} &|[dot,label=below:{(3,1)}] (p8)| {} & |[dot,label=below:{(1,4)}] (p9)| {} & \\
};
\draw (p1) edge node[above]{\scriptsize{$b$}} (p2);
\draw (p2) edge node[above]{\scriptsize{$d$}} (p3);
\draw (p3) edge node[above]{\scriptsize{$f$}} (p4);
\draw (p4) edge node[above]{\scriptsize{$h$}} (p5);
\draw (p1) edge node[left]{\scriptsize{$a$}} (p6);
\draw (p2) edge node[left]{\scriptsize{$c$}} (p7);
\draw (p3) edge node[left]{\scriptsize{$e$}} (p8);
\draw (p4) edge node[left]{\scriptsize{$g$}} (p9);
\end{tikzpicture}&
\begin{tikzpicture}[
dot/.style={shape=circle,fill=black,minimum size=0pt,
	inner sep=0pt,outer sep=0pt},
]
\matrix[matrix of nodes, column sep=3ex, row sep=4ex] {
	|[dot,label=above:C] (p1)| {} & |[dot,label=above:P] (p2)| {} & |[dot,label=above:C] (p3)| {} & |[dot,label=above:P] (p4)| {} & |[dot, label=right:{(6,3)}] (p5)| {}\\[1cm]
	|[dot,label=below:{(4,1)}] (p6)| {}&|[dot,label=below:{(1,2)}] (p7)| {} &|[dot,label=below:{(3,1)}] (p8)| {} & |[dot,label=below:{(1,4)}] (p9)| {} & \\
};
\draw (p1) edge node[above]{\scriptsize{$b$}} (p2);
\draw (p2) edge node[above]{\scriptsize{$d$}} (p3);
\draw (p3) edge node[above]{\scriptsize{$f$}} (p4);
\draw (p4) edge node[above]{\scriptsize{$h$}} (p5);
\draw (p1) edge node[left]{\scriptsize{$a$}} (p6);
\draw (p2) edge node[left]{\scriptsize{$c$}} (p7);
\draw (p3) edge node[left]{\scriptsize{$e$}} (p8);
\draw (p4) edge node[left]{\scriptsize{$g$}} (p9);
\end{tikzpicture}\\
Game 3&Game 4
\end{tabular}

\caption[]{Collection of the main games used in the experiment. The ordered pairs at the leaves represent payoffs for the computer ($C$) and the participant ($P$), respectively.\up} 
\label{fig:maingames}
\end{figure*}

\begin{figure*}[t]
\begin{center}
\begin{tabular}{cc}
\begin{tikzpicture}[
dot/.style={shape=circle,fill=black,minimum size=0pt,
	inner sep=0pt,outer sep=0pt},
]
\matrix[matrix of nodes, column sep=3ex, row sep=4ex] {
	 |[dot,label=above:P] (p2)| {} & |[dot,label=above:C] (p3)| {} & |[dot,label=above:P] (p4)| {} & |[dot, label=right:{(6,3)}] (p5)| {}\\[1cm]
	|[dot,label=below:{(1,2)}] (p7)| {} &|[dot,label=below:{(3,1)}] (p8)| {} & |[dot,label=below:{(1,4)}] (p9)| {} & \\
};
\draw (p2) edge node[above]{\scriptsize{$d$}} (p3);
\draw (p3) edge node[above]{\scriptsize{$f$}} (p4);
\draw (p4) edge node[above]{\scriptsize{$h$}} (p5);
\draw (p2) edge node[left]{\scriptsize{$c$}} (p7);
\draw (p3) edge node[left]{\scriptsize{$e$}} (p8);
\draw (p4) edge node[left]{\scriptsize{$g$}} (p9);
\end{tikzpicture}&
\begin{tikzpicture}[
dot/.style={shape=circle,fill=black,minimum size=0pt,
	inner sep=0pt,outer sep=0pt},
]
\matrix[matrix of nodes, column sep=3ex, row sep=4ex] {
	|[dot,label=above:P] (p2)| {} & |[dot,label=above:C] (p3)| {} & |[dot,label=above:P] (p4)| {} & |[dot, label=right:{(6,3)}] (p5)| {}\\[1cm]
	|[dot,label=below:{(1,2)}] (p7)| {} &|[dot,label=below:{(3,1)}] (p8)| {} & |[dot,label=below:{(1,4)}] (p9)| {} & \\
};
\draw (p2) edge node[above]{\scriptsize{$d$}} (p3);
\draw (p3) edge node[above]{\scriptsize{$f$}} (p4);
\draw (p4) edge node[above]{\scriptsize{$h$}} (p5);
\draw (p2) edge node[left]{\scriptsize{$c$}} (p7);
\draw (p3) edge node[left]{\scriptsize{$e$}} (p8);
\draw (p4) edge node[left]{\scriptsize{$g$}} (p9);
\end{tikzpicture}\\
Game $1'$&Game $3'$
\end{tabular}
\end{center}
\caption[]{Truncated versions of Game 1 and Game 3.\up\up }
\label{fig:auxgames}
\end{figure*}
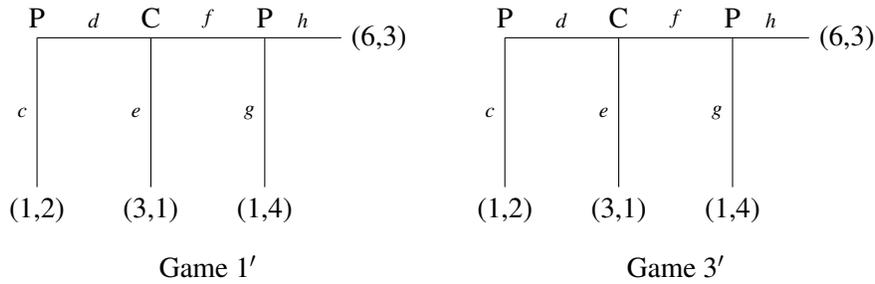

To get more conclusive evidence for FI reasoning on the part of the participants, we designed the current experiment, where the participants once again played 8 rounds of turn-based games, repeating in each round a set of 6 games as earlier. The main difference from the earlier experiment is in the payoff structures of the 6 games used for the experiment. We revised the payoff structures so as to minimize the probable cardinal effects 
reported in~\cite{ghv14}. The main features of the new payoff structures (cf. Figures \ref{fig:maingames} and  \ref{fig:auxgames}) are as follows:

\noindent - Game 1 differs from Game 2 only in the payoff of the Computer ($C$) following action $a$ (ending the game right away). As a result, Game $1'$ is the truncation of Game 1 as well as Game 2. This means that participants who get to play in Game 1 and Game 2 face exactly the same continuation game (Game $1'$). We assumed that these new payoffs would enable an unobstructed comparison between the participants' behavior across these games (i.e. without confounding effects of cardinal differences in payoffs, which may have interfered in the previous experiment~\cite{ghv14}). In~\cite{ghv14}, Game 1 and Game 2 had the same tree structures as currently, but the payoffs of player $C$ following $a$ and $e$ were interchanged, so that Game $1'$ was not a truncation of Game 2,   hence some cardinal effect may have been instrumental in participants' choices at the nodes of these games.

\noindent - The same holds good for Game 3 and Game 4, respectively.

\noindent - Game 1 differs from Game 3 only in   $P$'s payoff following $h$ at the very end. One can compare Game 2 and Game 4 in a similar manner. Previously, the payoff structure of Game 1 differed from that of Game 3, and the same for Game 2 and Game 4, respectively.

\noindent - We removed all the zero payoffs from these games, which some participants experienced as particularly ``bad" in the previous experiment (according to their verbal reports).

\noindent - At $C$'s second chance to move, exiting by playing $e$ guarantees $C$ a payoff of 3, which is only slightly smaller than the expected payoff 3.5 of a fifty-fifty lottery between $g$ (with which $C$ gets 1) and $h$ (with which $C$ gets 6).
This means that if at an earlier node, $P$ believes that $C$ is ``confused" (because $C$ has just deviated from its BI behavior), by attributing to $C$ a fifty-fifty belief on $P$'s future behavior at the last node (if reached),  $P$ cannot conclude what $C$ will do next (choose $e$ or $f$) if $P$ believes that $C$ is mildly risk-averse, because the precise level of $C$'s presumed risk aversion would determine whether a sure payoff of 3 is preferable or not to a fifty-fifty lottery between 1 and 6 in $C$'s eyes.

Furthermore, if due to the above, participant $P$ has a 50-50 belief on $C$'s future choice between $e$ or $f$ and $P$ is herself mildly risk-averse, cardinal considerations on their own will not give her a clear guidance what to do: exiting by playing $c$ will guarantee the payoff 2, while continuing by playing $d$ yields a fifty-fifty lottery between 1 (if $C$ chooses $e$) or 4 (if $C$ chooses $f$), with a slightly higher expected payoff of 2.5.

Thus, participants who are unsure how to interpret $C$'s initial deviation from BI will neither be able to come up with an easy forecast regarding $C$'s future behavior on the basis of cardinal payoff considerations, nor will they have a strong preference for $c$ or $d$ on the basis of cardinal payoff considerations---which were a confounding potential explanation for participants' behavior in the previous experiment, which we would like to circumvent.

To get some idea about BI and FI strategies in the current experimental games, let us now concentrate on  Game 1 (a variant of Reny's game~\cite{reny92}). (For the remaining games, the reader may adapt the reasoning strategies discussed in~\cite{ghv14}.) In Game 1, the unique Backward Induction (BI) strategy for player $C$ is $a;e$, while for player $P$ it is $c;g$. In case the last decision node of the game is reached, player $P$ will play $g$ (which will give $P$ a better payoff at that node) yielding 1 for $C$. Thus, in the previous node, if reached, $C$ will play $e$ to be better off. Continuing like this from the end to the start of the game by BI reasoning, it can be inferred that whoever is the current player will choose to end the game immediately. 

Forward induction, in contrast, would proceed as follows. Among the two strategies of player $C$ that are compatible with reaching the first decision node of player $P$, namely $b; e$ and $b;f$, only the latter is rational for player $C$. This is because   $b;e$ is dominated by $a;e$, while $b;f$ is optimal for player $C$ if she believes that player $P$ will play $d;h$ with a high enough probability. Attributing to player $C$ the strategy $b;f$ is thus player $P$'s best way to rationalize player $C$'s choice of $b$, and in reply, $d;g$ is player $P$'s best response to $b;f$. Thus, the unique Extensive-Form Rationalizable (EFR) strategy of $P$ is $d;g$, which is distinct from her BI strategy $c;g$. Nevertheless, player $C$'s best response to $d;g$ is $a;e$, which is therefore player $C$'s EFR strategy. Hence the EFR outcome of the game (with the EFR strategies $a;e$ and $d;g$) is identical to the BI outcome. A summary of the strategies in the games of Figures \ref{fig:maingames} and  \ref{fig:auxgames} is given in Table 1.

\begin{table}
\begin{center}
\begin{tabular}{ || l | l | l || }
\hline\hline
{\scriptsize\bf Games | Strategies} & {\scriptsize BI strategy} & {\scriptsize EFR strategy} \\ 
\hline
{\scriptsize Game 1} & {\scriptsize C: $a;e$} & {\scriptsize C: $a;e$} \\
				    & {\scriptsize P: $c;g$} & {\scriptsize P: $d;g$} \\	
\hline
{\scriptsize Game 2} & {\scriptsize C: $a;e$} & {\scriptsize C: $a;e$} \\
				    & {\scriptsize P: $c;g$} & {\scriptsize P: $c;g$} \\
\hline
{\scriptsize Game 3} & {\scriptsize C: $a;e, b;e, a;f, b;f$} & {\scriptsize C: $a;e, a;f, b;f$} \\
				    & {\scriptsize P: $c;g, d;g, c;h, d;h$} & {\scriptsize P: $d;g, d;h$} \\	
\hline
{\scriptsize Game 4} & {\scriptsize C: $a;e, b;e, a;f, b;f$} & {\scriptsize C: $a;e, b;e, a;f, b;f$} \\
				    & {\scriptsize P: $c;g, d;g, c;h, d;h$} & {\scriptsize P: $c;g, d;g, c;h, d;h$} \\
\hline
{\scriptsize Game $1'$} & {\scriptsize C: $e$} & {\scriptsize C: $e$} \\
				    & {\scriptsize P: $c;g$} & {\scriptsize P: $c;g$} \\	
\hline
{\scriptsize Game $3'$} & {\scriptsize C: $e, f$} & {\scriptsize C: $e, f$} \\
				    & {\scriptsize P: $c;g, d;g, c;h, d;h$} & {\scriptsize P: $c;g, d;g, c;h, d;h$} \\
\hline\hline
\end{tabular}
\label{game-summary}
\end{center}
\caption{BI and EFR (FI) strategies for the 6 experimental games in Figures 1 and 2\up\up}\up\up
\end{table}

We now discuss the experimental procedure, which is almost the same as reported in~\cite{ghv14}. A group of 50 Bachelor's and Master's students from different disciplines   participated in the experiment. As in~\cite{ghv14}, the participants had little or no knowledge of game theory, so as to ensure that neither backward induction nor forward induction reasoning was already known to them. 96\% of the participants were aged between 18 and 32. 48\% of the participants were female, 52\% of the participants were male. All experiments were conducted at the Institute of Artificial Intelligence 
at the University of Groningen, in the ALICE Lab. 

\begin{figure}[h]
\begin{center}
\includegraphics[width=.5\textwidth]{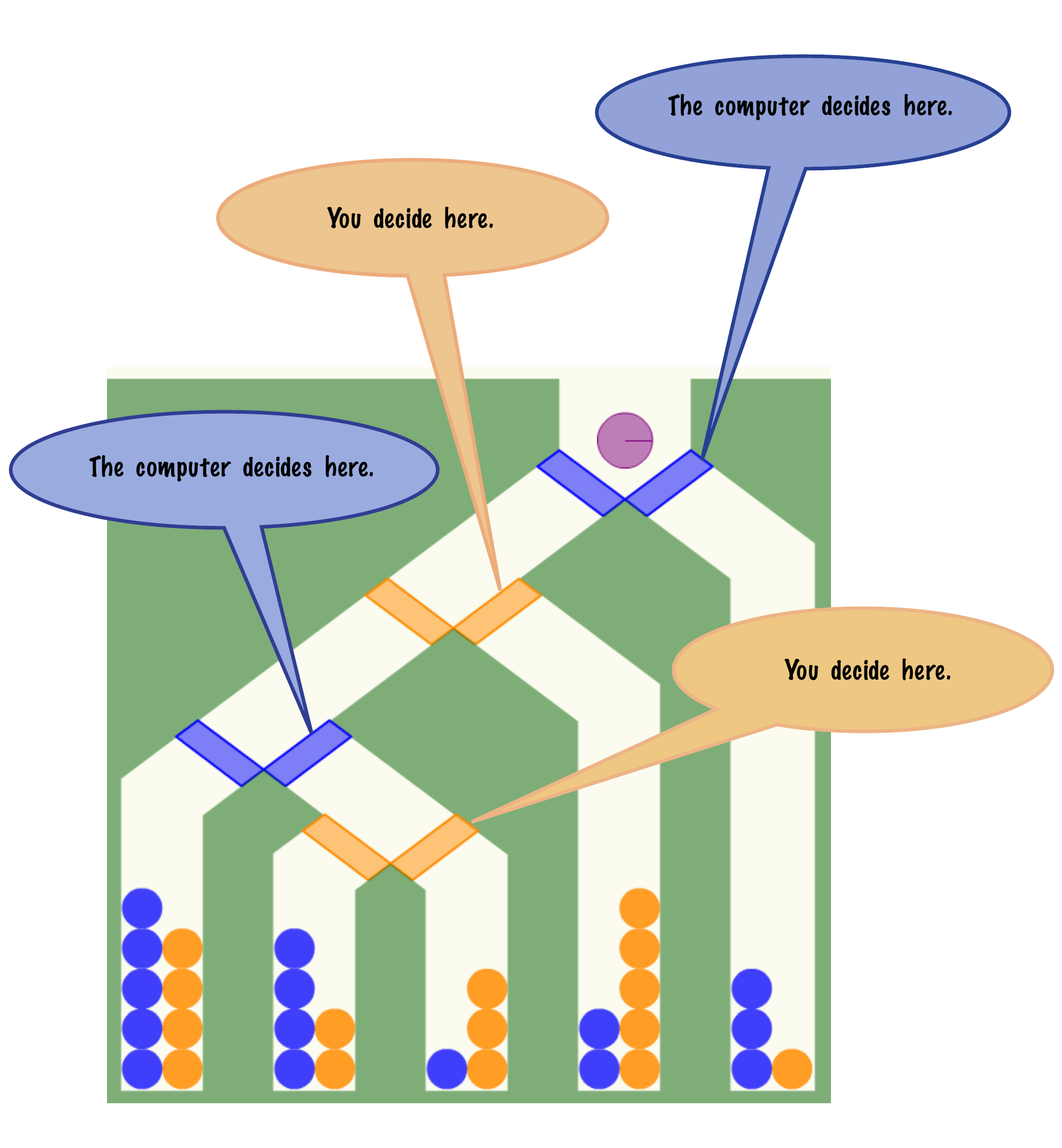} \up\up\up
\end{center}
\caption[]{Graphical interface for the participants.\up\up} 
\label{fig:interface}
\end{figure}

The participants played the finite perfect-information games in a graphical interface on the computer screen (cf. Figure \ref{fig:interface}).  In each game, a marble was about to drop, and both the participant and the computer determined its path: The participant controlled the orange trapdoors, and the computer controlled the blue trapdoors. The participant's (computer's) goal was that the marble should drop into the bin with as many orange (blue) marbles as possible. 

At the start of the experiment, participants received verbal instructions (based on an instruction sheet, see Appendix A) regarding the experiment. These instructions emphasized two facts in particular. Participants were advised to regard each game  as if it were played against a new opponent. Furthermore, participants were   notified about the fact that at the start of each new game, the computer had its strategy planned out, and that strategy was a best response to a strategy that the computer assumed the participant would play.

The participants first played 14 practice games of increasing difficulty so as to get acquainted with the game setting. The experimental phase consisted of 8 rounds. In each round, the participants played the six games that were described above against a computer opponent. The order in which these 6 games were played in each round was randomized. In each new round, the graphical representation of each game was altered, so as to minimize the possibility for participants to recognize the games they played in earlier rounds.

\begin{figure}[h]
\begin{center}
\includegraphics[width=.8\textwidth]{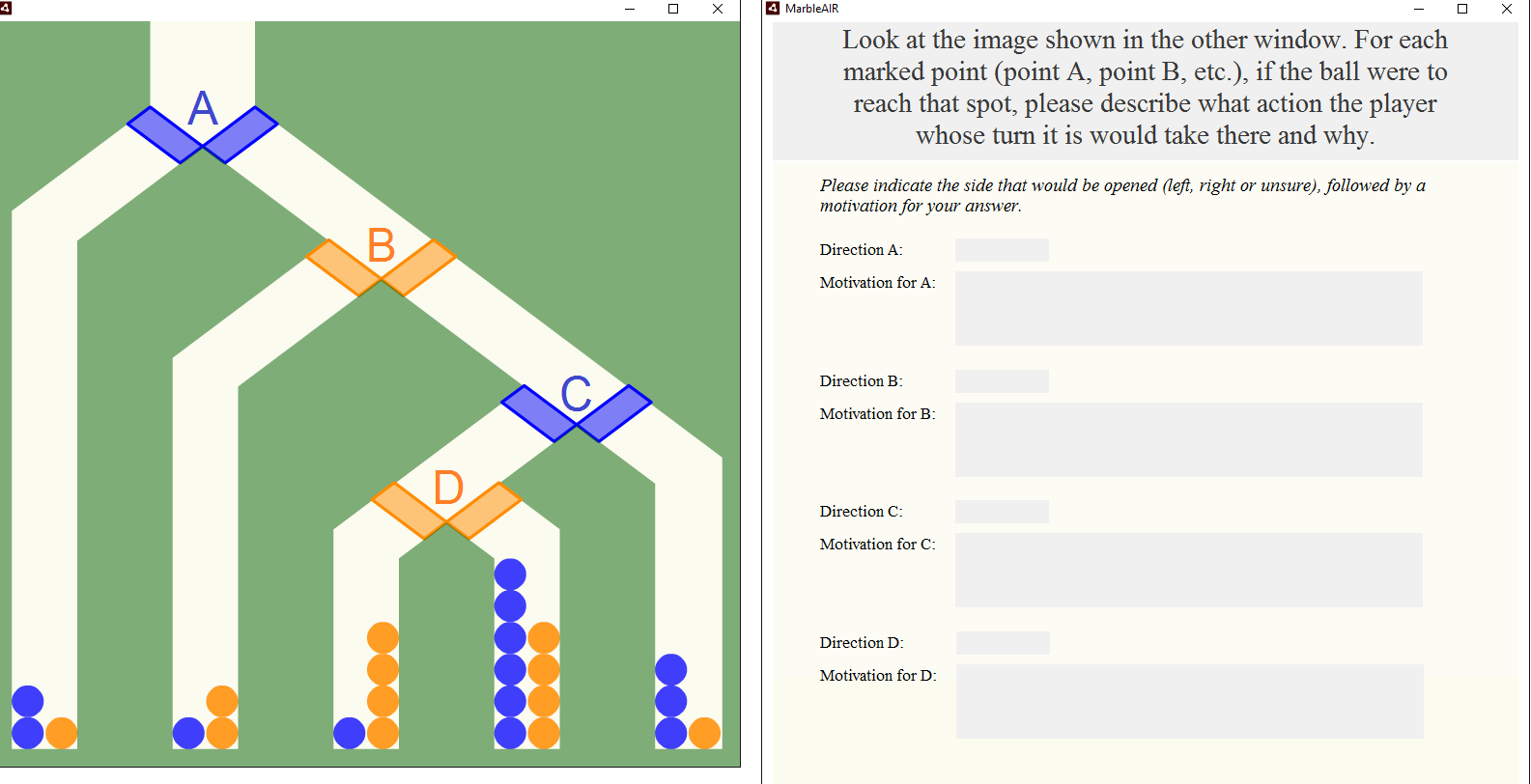}\up\up
\end{center}
\caption[]{Questionnaire based on a representation of Game 4.\up}
\label{fig:finalquestion}
\end{figure}

The participants were randomly divided into two groups: Group A and Group B, each consisting of 25 persons. While playing in certain rounds of a game, as they were about to make their choices at their first decision node, Group A participants were asked a multiple-choice question as follows. (i) For games 1-4: ``The computer just chose to go [direction computer just chose]. If you choose to go [direction corresponding to playing $d$], what do you think the computer would do next?'', or, (ii) For games $1', 3'$: ``It's your turn. If you choose to go [direction corresponding to playing $d$], what do you think the computer would do next?''. Three options were given regarding the likely choice of the computer: ``I think the computer would most likely open the left side" or `` I think the computer would most likely open the right side" or ``Both answers seem equally likely".  The first two answers translated to the moves $e$ or $f$ of the computer, respectively. In case of the third answer, we assumed that  the participant was undecided regarding the computer's next choice. 

For the Group B participants, questions were asked in certain rounds at the end of a game: (i) For games 1-4: ``The computer first chose to go [direction computer chose at its first decision point]. When you made your first choice, what did you think the computer would do next if you chose to go [direction corresponding to playing $d$]?'', or, (ii) For games $1', 3'$: ``When you made your first choice, what did you think the computer would do next if you chose to go [direction corresponding to playing $d$]?''. The answer choices were the same as above. 

At the end of the experiment, the participants were presented with two questionnaires, both based on an image of one of the experimental games 
(cf. Figure \ref{fig:finalquestion}, based on a representation of Game 4). All participants were asked these questions based on the same images. Finally, they  were paid according to the marbles they earned at one of the experimental games, selected randomly for each participant by a mechanism proposed by Allais~\cite{allais53}, and supported as incentive-compatible by Azrieli et al.~\cite{healy12}. The reward varied from 3.75 euros to 15 euros depending on the number of marbles (1--4) that the participant won in the randomly selected game.


\section{Results and analysis}\label{sec:results}


 We did not find any significant difference in  behavior between participants in groups A and B. Instead, we found strong evidence that participants were equally likely to choose to continue (playing $d$) at the first decision node (Bayes Factor = 0.056). Consequently, we analyze the data of all 50 participants together.
 
 \begin{figure}[p]
\begin{center}
\begin{tabular}{cc}
\includegraphics[width=.45\textwidth]{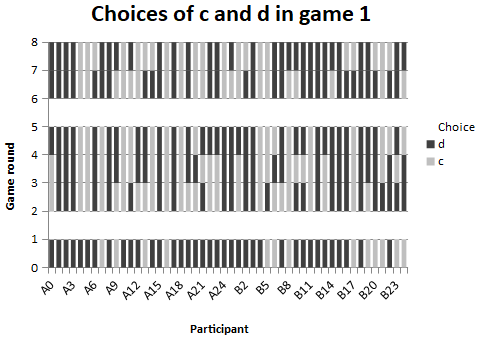} & \includegraphics[width=.45\textwidth]{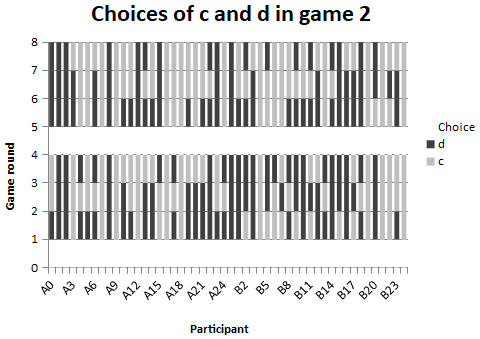}\up\up\up\up\up  \up\up\up\up\\ 
\includegraphics[width=.45\textwidth]{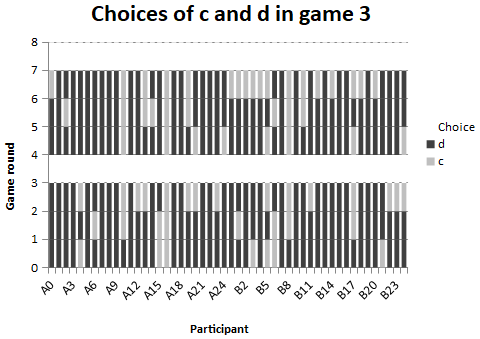} & \includegraphics[width=.45\textwidth]{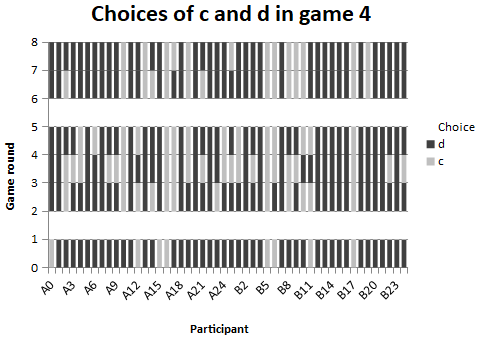}\up\up\up\up\up  \up\up\up\up\\ 
\includegraphics[width=.45\textwidth]{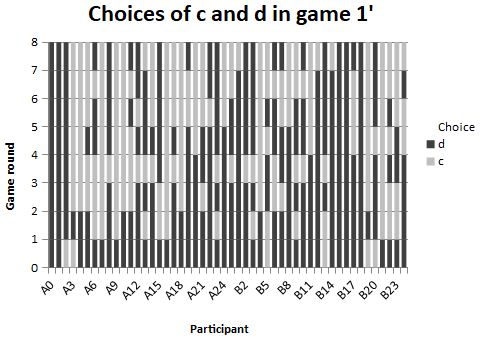} & 
\includegraphics[width=.45\textwidth]{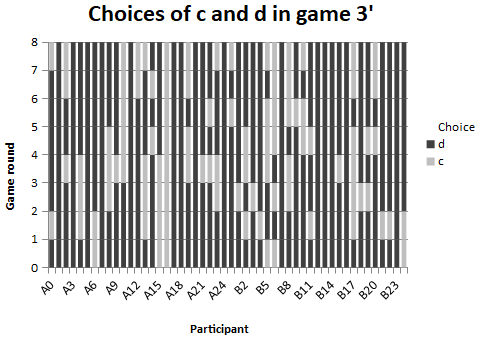}\up \up \up\up\up  \up\up\up\up\\
\end{tabular}
\end{center}
\caption[]{Sequence of choices (across the 8 repetitions of each game) at the first decision node of games 1, 2, 3, 4, $1', 3'$, per participant (named A1 \ldots A25, B1  \ldots B25). The {\em dark grey} color corresponds to the rounds the participant played the move $d$, and the {\em light grey} color corresponds to the rounds the participant played move $c$, whenever the participant's first decision node was reached. Note that white horizontal bands correspond to rounds in which the computer took option $a$, thereby ending the game.\up\up\up} 
\label{fig:choices1}
\end{figure}

\subsection{Aggregate results: Is there a  forward induction trend at participants' first node?}

Our main hypothesis was that in case the participants played $c$ more in Game 2 than in Game 1, and played $c$ more in Game 4 than in Game 3, then we could safely assume that the participants' behavior showed some inclination towards FI reasoning. To test this hypothesis, we first perform a mixed effects logistic regression on participant choices in Game 1 and Game 2. In this model, we determine whether the probability that a participant chooses $c$ in a given game depends on whether the game is an instance of Game 1 or  
an instance of Game 2. To account for individual differences, participants are treated as random effects. The results show that participants were indeed significantly more likely to play $c$ in Game 2 than in Game 1 ($p < 0.01$). A second logistic regression for Game 3 and Game 4 shows that that participants were also significantly more likely to play $c$ in Game 4 than in Game 3 ($p < 0.02$). These results suggest that in the aggregate, participants' behavior shows inclination towards FI reasoning.

However, on a visual inspection of the graphs showing the choices of the participants at their first decision nodes (cf. Figure \ref{fig:choices1}) we found the following: Comparing Game 1 to Game 2, only 14 participants out of 50 played $c$ in Game 2 somewhat more than they played $c$ in Game 1 (meaning that the number of times such a participant played $c$ in the rounds of Game 2 minus those he played $c$ in the rounds of Game 1 is more than 1). In contrast, 5 participants played $c$ in Game 2 somewhat less than they played $c$ in Game 1. For the remaining 31 participants, there was not much difference in  their playing $c$ between Games 1 and 2. Comparing Game 3 to Game 4, only 10 participants out of 50 played $c$ somewhat more in Game 4 than they did in Game 3 (with no reverse cases), and for the remaining 40 participants there was not much difference between their playing $c$ across Games 3 and 4. We finally note that only 13 participants out of 50 played $c$ in more than half of the 8 rounds of Game $1'$. Interestingly, all these trends in participants' choices at their first decision points indicated a very low extent of game-theoretic strategic reasoning overall---be it BI or FI reasoning.

\begin{figure}[h]
\begin{center}
\begin{tabular}{ll}
\includegraphics[width=.45\textwidth]{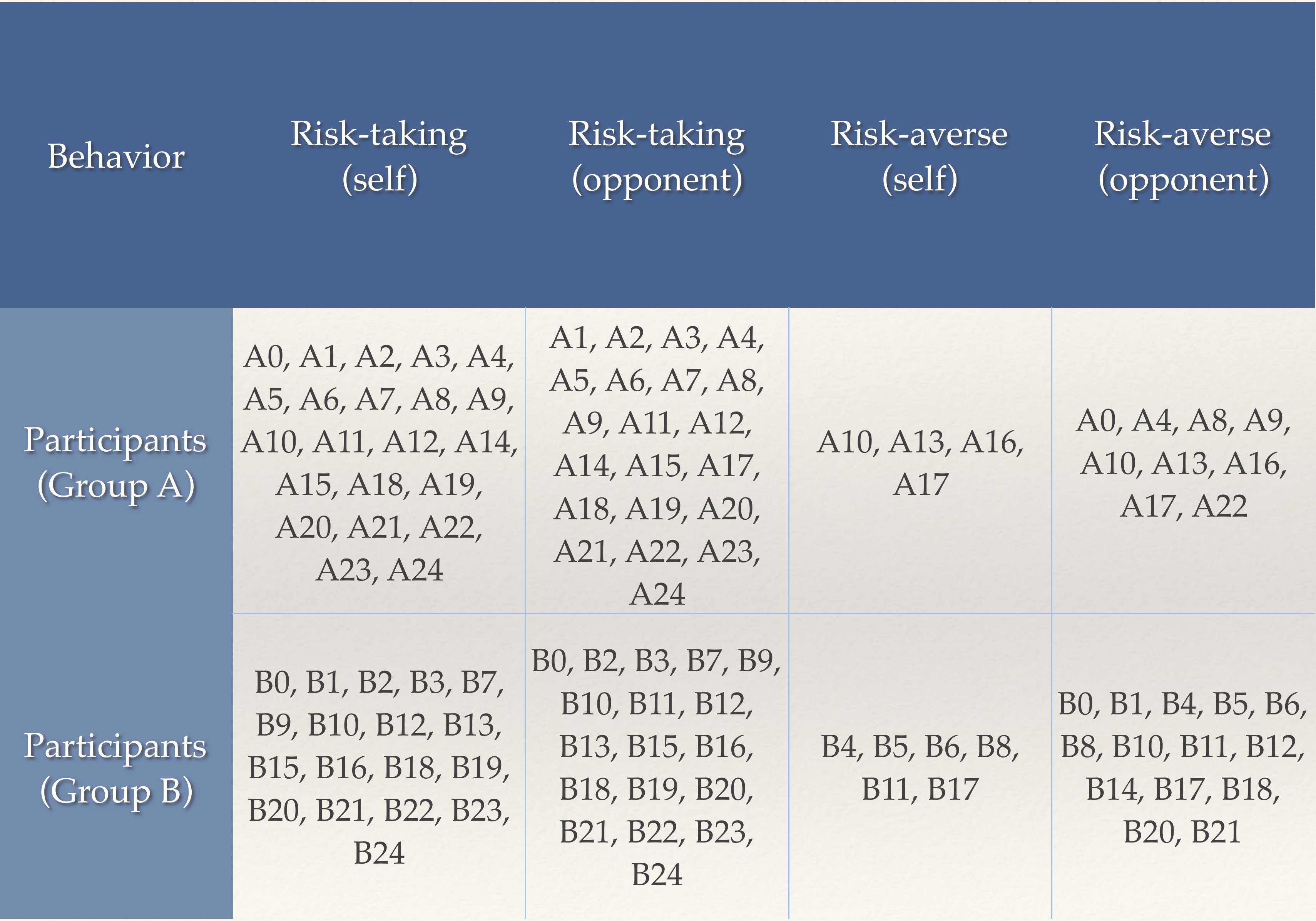} & \includegraphics[width=.41\textwidth]{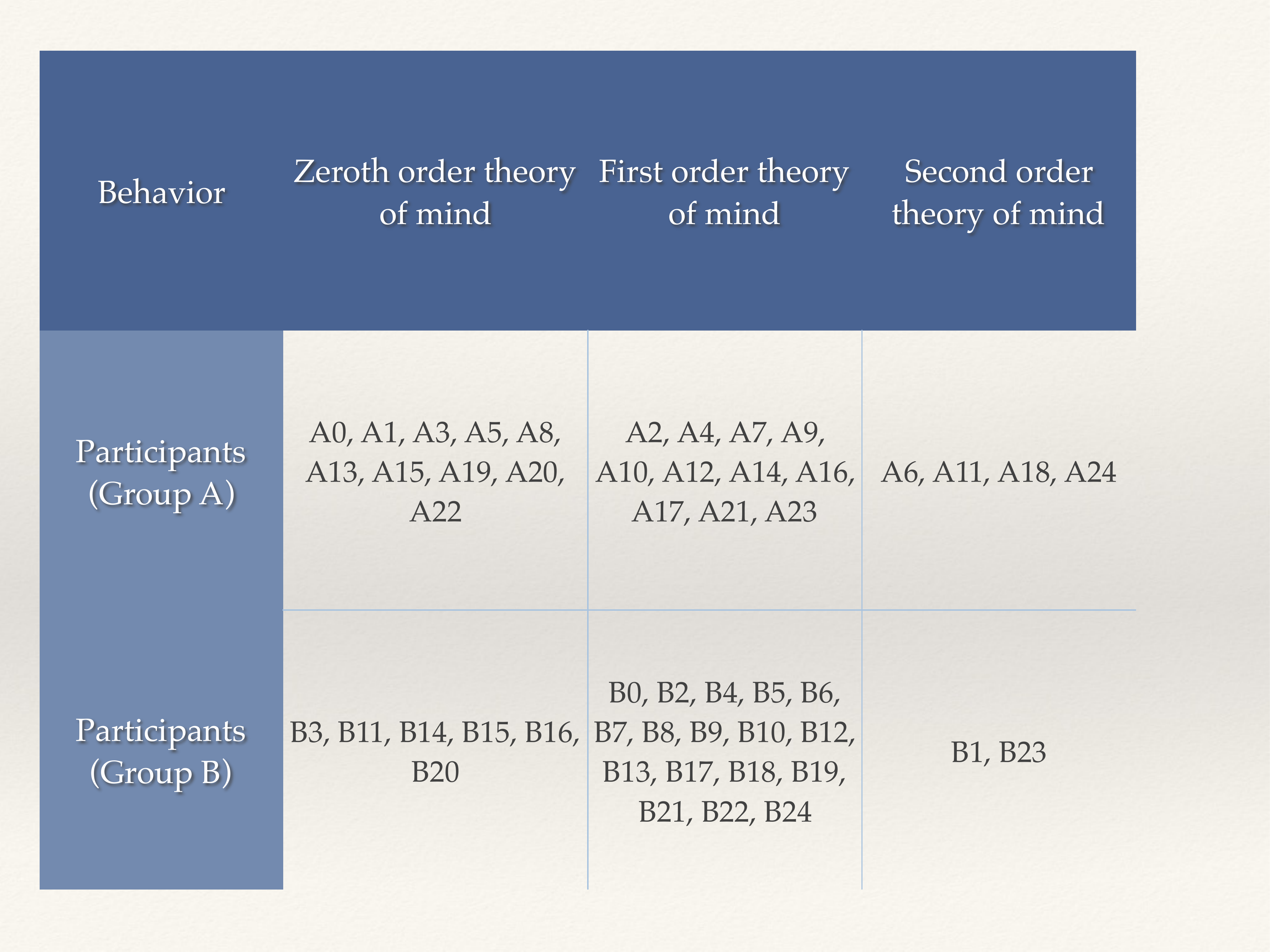}\\ 
\end{tabular}
 \includegraphics[width=.45\textwidth]{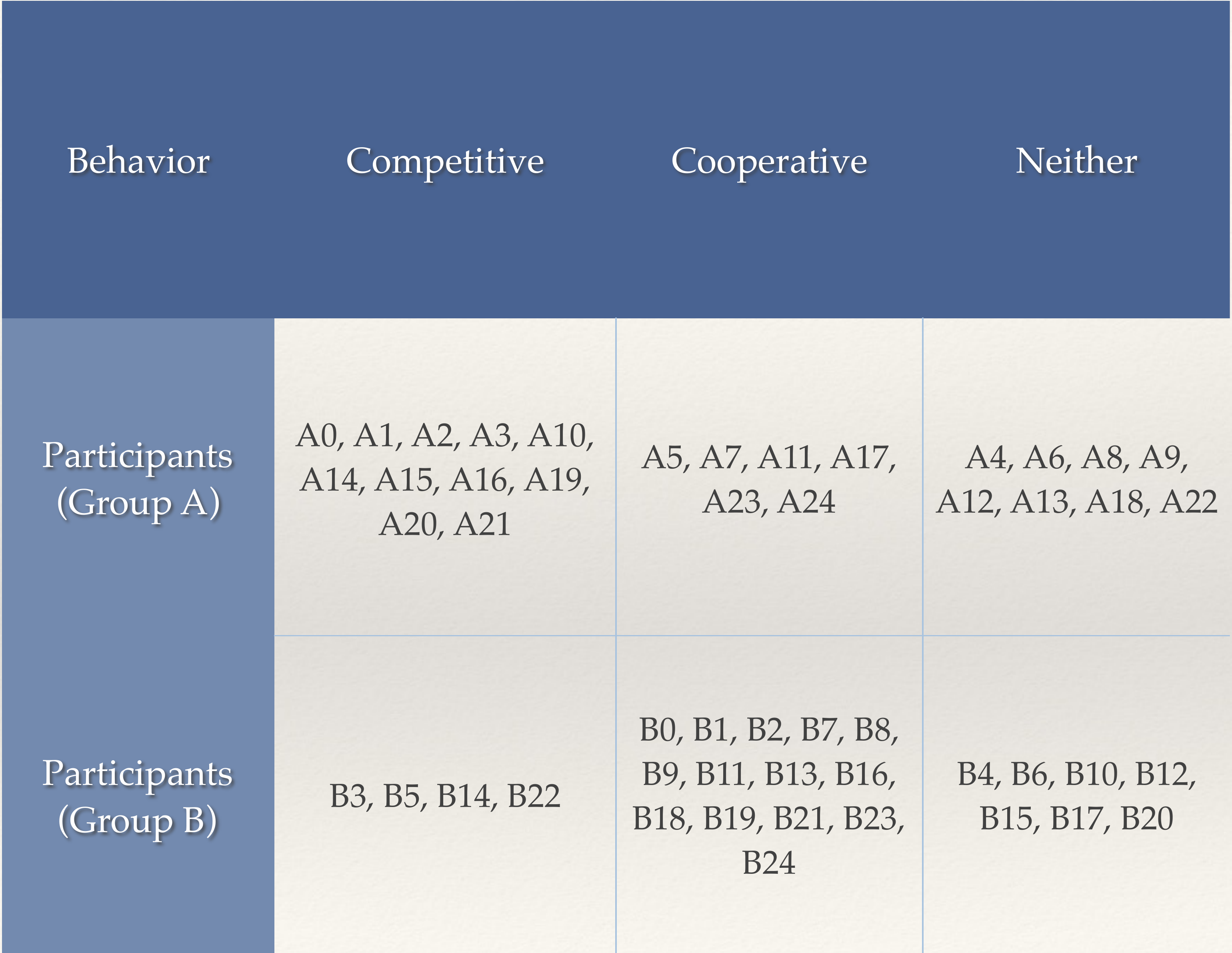}\up\up
\end{center}
\caption[]{Reasoning behind participant behaviour. Answers to final question classified on three dimensions: attitude towards risk (left); explicit order of theory of mind (middle) and  level of cooperation (right) \up}
\label{fig:finalanswer}
\end{figure}


We should note here that in the current experiment, in all six games, the computer would gain 6 points if the game ended up at the rightmost leaf, compared to only 4 points at the rightmost leaves in the games of~\cite{ghv14}. This higher pay-off appears to have motivated the participants to think that the computer would risk for 6 and would continue in the game.

Given these conflicting findings of the logistic regression and the visual inspection of the choice graphs, it is pertinent to ask: ``How exactly do participants reason while playing these games?" In what follows, we attempt to answer this question. 

\subsection{Dividing participants into groups: decisions and answers about opponent}

One way to divide the participants into groups is to analyze their answers to the questionnaire presented in Figure \ref{fig:finalquestion}. According to their answers, we classify the participants along the following three dimensions (see Figure \ref{fig:finalanswer}): (i) {\em risk-taking/risk-averse} considerations for self and opponents, (ii) explicit order of {\em theory of mind}, which has been shown to be important in strategic reasoning in dynamic games~\cite{meijering2010,meijering2011}, and (iii) {\em competitive/cooperative} considerations. To come up with our classification, we considered the decisions at each node of the game in Figure \ref{fig:finalquestion} in the questionnaire and the participants' explanations of reasoning behind their decisions.

As an example, consider the following sample answers of the participants A7 and B0 to the questionnaire given in Figure \ref{fig:finalquestion}:\footnote{The full answers of all participants' answers to the final questionnaire will be provided on~\url{http://www.ai.rug.nl/SocialCognition/experiments/}.}

\medskip

\noindent {\em Participant A7:}\\
\noindent
{\bf Direction A}: right

\noindent
{\em Motivation:}\ B would probably open right, at which point C has the chance to get at least 3 marbles.

\noindent
{\bf Direction B}: right

\noindent
{\em Motivation:}\ If C opens left, B has a 100\% chance of getting 4 marbles. The chance that C opens left is big enough, since there are 6 marbles in D-right. 

\noindent
{\bf Direction C}: left

\noindent
{\em Motivation:}\ Since the direction for D doesn't matter, and the amount of marbles of your opponent doesn't matter, the chance is quite big that D will choose right, giving you 6 marbles.

\noindent
{\bf Direction D}: right

\noindent
{\em Motivation:}\ I would pick right. Even against the computer it felt wrong to pick left (since the opponent's score doesn't matter for my own score).

\medskip

\noindent {\em Participant B0:}

\noindent
{\bf Direction A}: right

\noindent
{\em Motivation:}\ The player would hope for the largest amount of points he could get. 2 would not be enough when there is the chance to get three or six.

\noindent
{\bf Direction B}: right

\noindent
{\em Motivation:}\ The player would chose right because there his/her chances are good that he/she will end up with 4 balls. But there is the risk of having only one ball

\noindent
{\bf Direction C}: right

\noindent
{\em Motivation:}\ If the player decides to go left, he will let the other person choose how many balls he will end up with. This is a high risk since it is possible to get just one ball. Therefore the player would take the right path were he has the mean of 3.

\noindent
{\bf Direction D}: right

\noindent
{\em Motivation:}\  The player would chose right since both boxes contain 4 balls. Further it is not important to let the other player have less balls than you. Therefore I would say the player is nice and let the other person have the full amount of balls.\\

\medskip

\noindent
Based on these answers, we came up with the classification shown in Figure \ref{fig:finalanswer}. Participant A7 considers the chance of getting more for both the players (risk-taker: self and opponent), takes into consideration how the opponent would play (first-order theory of mind), and does not want to wrong the opponent (cooperative). Participant B0 also considers the chance of getting more for both the players (risk-taker: self and opponent), in addition, considers safe play for the opponent (risk-averse: opponent), takes into consideration how the opponent would play (first-order theory of mind), and wants to be nice to the opponent (cooperative). Note that Directions A and C in Figure~\ref{fig:finalquestion} (blue trapdoors) are considered to be the computer's moves, and Directions B and D (orange trapdoors) are considered to be the participant's moves in the interpretation here.

\begin{figure}
\begin{center}
\includegraphics[width=.65\textwidth]{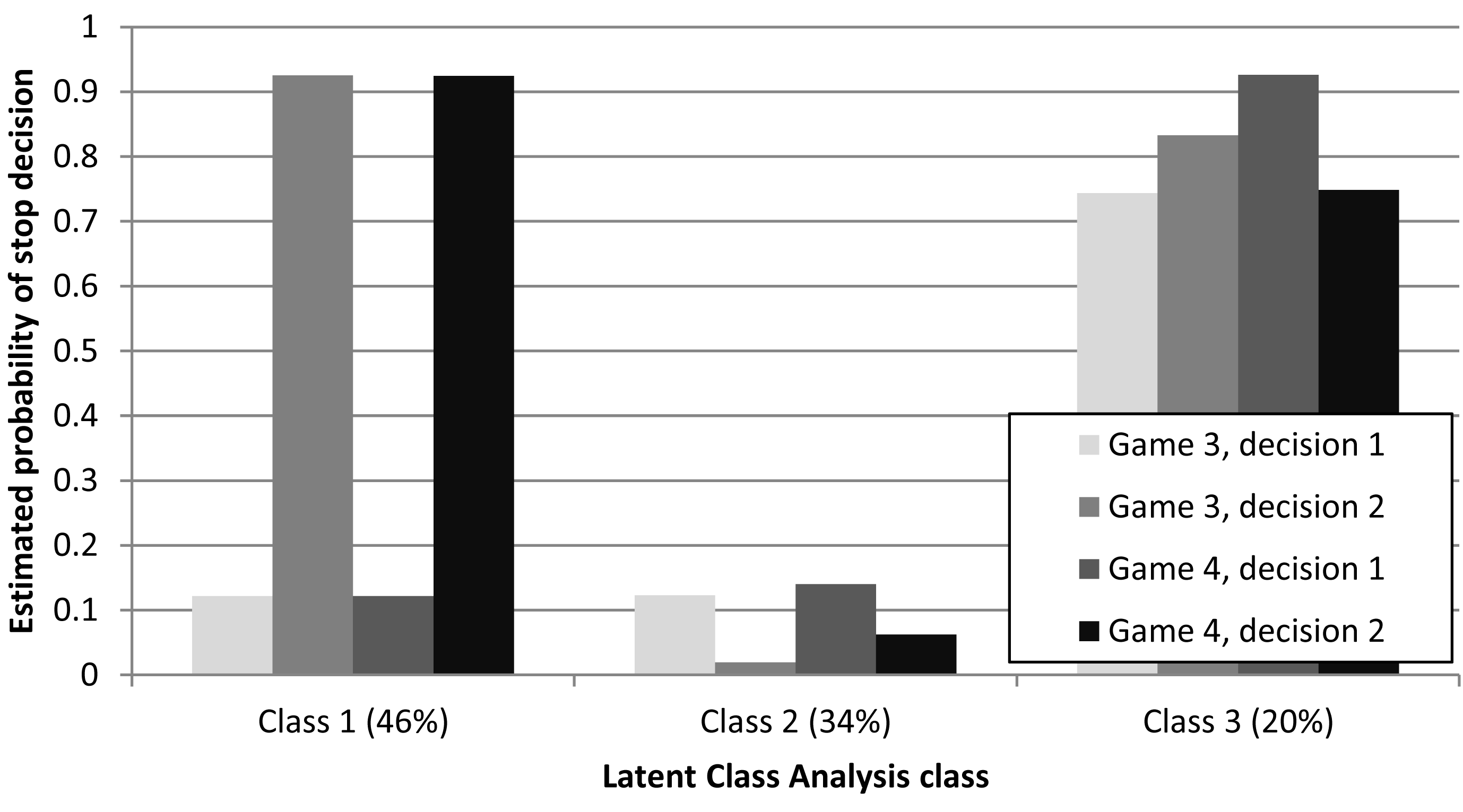}\up\up
\end{center}
\caption{Estimated probability of stop decision (i.e. $c$ or $g$) according to the latent class analysis of participant choices on Game 3 and Game 4.\up\up}
\label{fig:lca34}
\end{figure}

\subsection{Dividing the participants by latent class analysis according to their decisions}

To further analyze participant choices, we used latent class analysis (LCA, \cite{mccutcheon1987latent}), allowing us to divide participants into classes that exhibit similar patterns of behavior. In our model, the behavior of a participant is described by whether they choose $c$ or $d$ at their first decision point of a game, and whether they choose $g$ or $h$ at their second decision point. LCA determines what classes of behavior best describe the observed participant data, given the number $n$ of such classes. Statistical model selection tools can be used to select the number of classes that is most appropriate for the data.

Our aim of the latent class analysis is to describe the behavior of participants in both their decision points. However, of the 181 times (out of 300 times where the participants actually reached their first decision node) that a participant in Game 1 ended up at his or her second decision point, only 2 times participants chose $h$. In addition, no participant in Game 2 ever reached the second decision point---the computer was programmed to play $e$ whenever its second decision node was reached. For this reason, we performed a latent class analysis on the two decision points of Game 3 and Game 4 only. The Bayesian Information Criterion (BIC) favors an LCA model with three classes, which we describe here. The results of the estimation parameters of the LCA are summarized in Figure \ref{fig:lca34}. In this figure, estimated probabilities of choosing $c$ at the first decision point or $g$ at the second decision point are averaged over repeated encounters of the same game. According to the LCA, 46\% of the participants tend to choose to continue at the first decision point, but stop at the second decision point: Class 1. A further 34\% of the participants is represented by the second class, who prefer to continue at both decision points: Class 2. The final 20\% of the participants has a preference to stop at both decision points: Class 3.


Thus, about 80\% of the participants belonged to Class 1 or 2, who tend to continue at the first decision node, and the remaining 20\% tend to stop. Out of these 41 participants assigned to Class 1 or Class 2 by LCA, 37 were classified as risk-takers (self), 37 were classified as risk-takers (opponent), while only two participants were classified as neither type of risk-taking based on their final answers (see Figure \ref{fig:finalanswer}). Out of the 9 participants assigned to Class 3 by LCA, 7 were classified as risk-averse (self), 4 were classified as risk-averse (opponent), while only one participant did not mention any risk-aversion, once again, based on their final answers.

As depicted in Figure \ref{fig:maingames}, the final decision point for Player $P$ in Game 3 and Game 4 only influences the payoff for the computer player $C$. Participants that ended up at this decision point more often chose the competitive option (60\%), in which the opponent only received a payoff of 1, than the cooperative option (40\%), which yielded the opponent a payoff of 6 (Bayes Factor = 132.7663). In addition, of the 19 participants assigned to Class 2 by the LCA, who are therefore classified as being likely to choose the cooperative option at their second decision point, 15 were also classified as being cooperative based on their final answer (see Figure \ref{fig:finalanswer}). Moreover, of the 31 remaining participants, only 5 were classified as being cooperative based on their final answer. However, there was no clear mapping of classes 1 and 3 of the LCA results and the classification `competitive' and `neither cooperative nor competitive' based on the final answers.

\begin{figure}
\begin{center}
	\includegraphics[width=.65\textwidth]{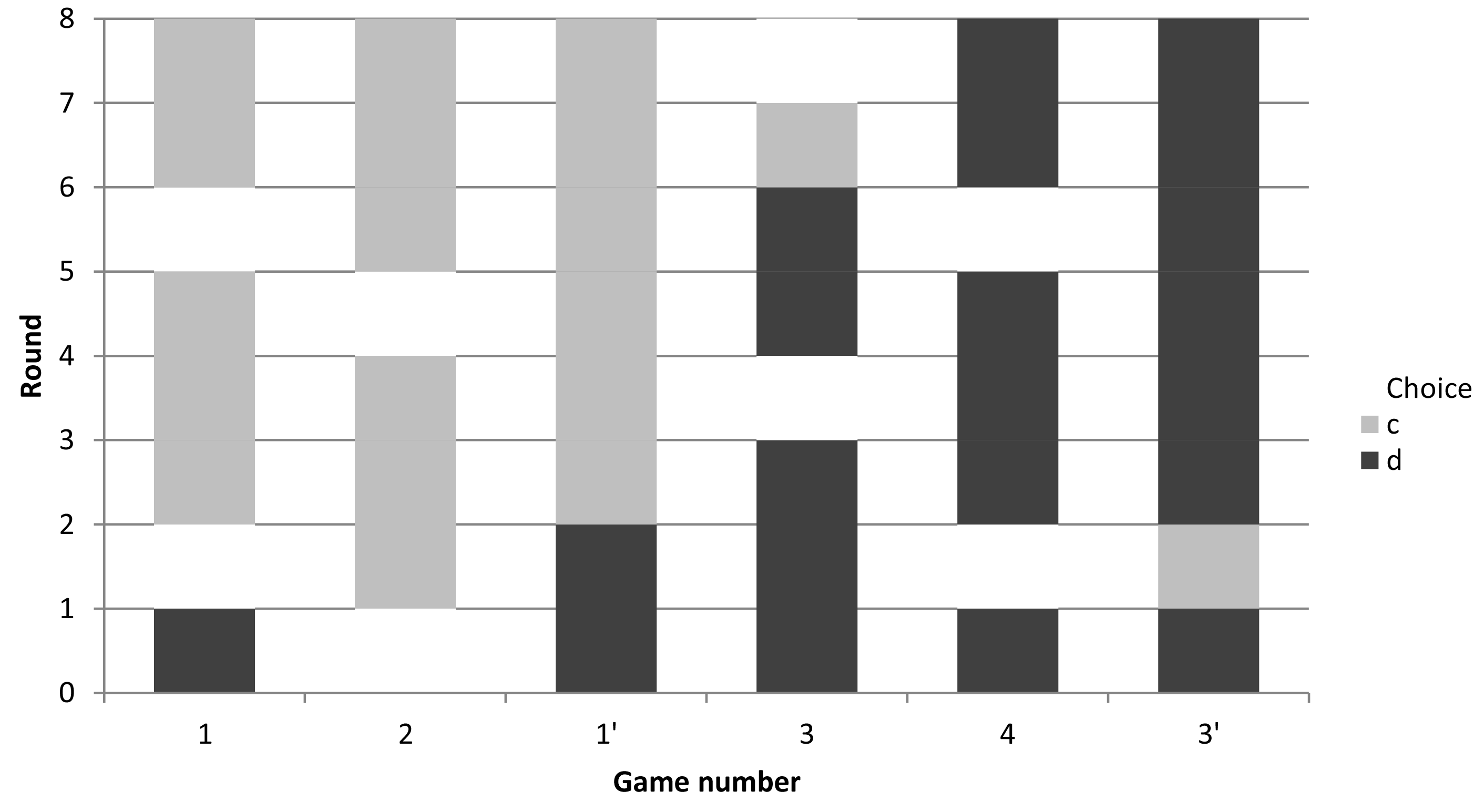}\up\up
	\end{center}
	\caption{Example of participant choices at the first decision point across all games. This participant, like many others, tended to choose $c$ at the first decision point for Game 1, Game 2, and Game $1'$, but tended to choose $d$ for Game 3, Game 4, and Game $4'$.\up\up}
	\label{fig:playergraph}
\end{figure}

\subsection{Individual participant decision patterns: differences and similarities}

Visual inspection of individual participant graphs shows a similar pattern. See Figure \ref{fig:playergraph} for an example of a graph of an individual participant's choices: In two of the rounds for games 1-4, selected randomly, the computer took the \emph{outside option}, that is, played $a$, and so the first decision node of the participant was not reached. It was noted that participants do differentiate between Game 1 and Game 2 on the one hand and Game 3 and Game 4 on the other hand: The participants played $c$ more often in Games 1 and 2, compared to what they played in Games 3 and 4. Further, even though participants' behavior differed between Game 1 and Game 2 to some extent (not exceedingly so), participants appeared to behave similarly in Game 3 and Game 4. For the truncated games, behavior in Game $1'$ appeared similar to Game 1, and behavior in Game 3' appeared similar to both Game 3 and Game 4.


\section{Discussion}\label{sec:related}

The results outlined in Section \ref{sec:results} indicate that while in the aggregate, participant behavior is indicative of forward reasoning, most participants do not appear to making use of either forward induction reasoning or backward induction reasoning. Apparently, game-theoretic rationality does not drive participants' choices in these turn-taking games. In this section, we take a look at some possible alternatives for the driving force behind people's choices in our experiment, in the light of the literature on human decision making and perspective taking in games.

\subsection{Level-$k$ reasoning}

In our experiment, we explicitly mention to the participants that the computer player acts based on its beliefs about future participant behavior, and that the computer player does not learn from the participant's previous actions. Participants are therefore playing a series of one-shot games against a computer opponent. In behavioral economics, the behavior of participants in such one-shot games has been successfully modeled through iterated best-response models such as level-$n$ theory \cite{stahl1995players,Bacharachstahl2000leveln,costa2001cognition,Nagel1995,bn08,Crawford2013}, cognitive hierarchies \cite{camerer2004cognitive}, quantal response equilibria \cite{mckelvey1995quantal}, and noisy introspection models \cite{goeree2004model}. 

Similar to the theory of mind classification we describe in Section \ref{sec:results}, a participant's level of reasoning sophistication is measured by the maximum number of steps of iterated reasoning the participant considers. 
According to cognitive hierarchy models \cite{camerer2004cognitive}, a naive level-0 reasoner does not reason strategically at all, but instead chooses randomly among all available options. In contrast, a level-1 reasoner believes that all other participants are level-0 reasoners, and selects the option that is a best response to that belief. In terms of our Marble Drop game, level-1 reasoners would therefore maximize future payoffs without considering the opponent's past behavior.
In the cognitive hierarchy model, a level-2 reasoner performs two steps of iterated reasoning. Such a reasoner believes that other individuals can be either level-0 reasoners that play randomly, or level-1 reasoners that maximize their expected payoff. In addition, level-2 reasoners form beliefs about the relative proportion of level-0 reasoners and level-1 reasoners.

Over a range of one-shot non-repeated games, cognitive hierarchy models estimate participants to be level 1.5 reasoners on average \cite{camerer2004cognitive,costa2006cognition}. In terms of theory of mind reasoning, the vast majority of participants in these one-shot non-repeated games reason at zero-order or first-order theory of mind. Although part of the participants in those experiments were found to use more than two steps of iterated reasoning, only few players were found to be well-described as higher-level agents \cite{wright2010beyond}. 

\subsection{Considering the opponent's perspective}

Kawagoe has recently applied a level $k$-analysis to centipede games to explain that participants' lack of  backward induction may be due to their believing that the opponent could have made an error or cannot apply backward induction for the number of steps required~\cite{Kawagoe2012}. Evans and Krueger have shown   in a very simple dynamic trust game of perfect information, in which participants need to open windows in order to inspect payoffs, quite a few participants do not choose to inspect all the opponent's payoffs but they do inspect all their own ones, thereby not applying even first-order theory of mind~\cite{Evans2011,Evans2014}. For higher orders of theory of mind, the situation seems even worse. It has also been shown that level $k$ reasoning for higher $k$ is correlated to general cognitive ability~\cite{Gill2016}. 

These results could make one quite pessimistic about the value and actual use of higher-level per\-spective-taking in games.
Is the situation really that bad? De Weerd and colleagues have shown on the basis of agent simulations that both in competitive situations and in repeated mixed-motive interactions, agents attain higher pay-offs if they apply second-order theory of mind than first-order theory of mind, which is in turn more beneficial than zero--order theory of mind~\cite{Weerd2013,Weerd2017}, so the application of theory of mind is beneficial. Also theoretically, higher-order theory of mind is required in dynamic perfect information games~\cite{Pacuit2015}. But do people really use it? De Weerd and colleagues~\cite{Weerd2017} showed that, by letting people play unknowingly against a second-order theory of mind agent in a repeated negotiation game, people are enticed to apply second-order theory of mind, which became much more prevalent than when the participants played against zero-order or first-order agents. 

In cognitive science, the application of perspective taking in turn-taking games of perfect information has been studied, especially on the basis of experiments in which the decision trees looked like our Game 1$'$ and 3$'$, but with a large variation of pay-off structures, not only those similar to centipede games. These various turn-taking games, when solved by participants without knowledge of the backward induction algorithm, do require second-order theory of mind: ``the opponent {\em  thinks} that at my final move, I {\em intend} to go down".
 Hedden and Zhang~\cite{hedden2002} showed that people have a hard time learning to apply second-order in these games (around 60 \% optimal decisions after many game items). Meijering and colleagues~\cite{meijering2010,meijering2011} proposed several supportive interventions that helped people to make much better decisions (up to around 90 \% optimal ones at the end of playing many experimental games with different payoff structures). 
 
Even though application of second-order theory of mind can apparently be supported and trained, Meijering and colleagues~\cite{meijering2014a} also showed on the basis of computational cognitive models that people do not start to apply higher orders of theory of mind spontaneously. Rather, people start to do this only when they receive negative feedback from the results of the games they play, for example, in the form of low pay-off, or the comment ``you could have chosen better''. They prefer reasoning that is ``as easy as possible, as complex as necessary". And even when they do produce the backward induction {\em outcome}, it appears by analyzing their eye movements and reaction times that their reasoning rather follows a stream of thinking starting from the root to its child nodes and then to the grandchildren in the decision tree, by ``forward reasoning plus backtracking", and giving special attention to decision points~\cite{meijering2012,Bergwerff2014}.

 \subsection*{Coming back to the experiment}
 
The experiments done by~\cite{meijering2010,meijering2011,meijering2012} concern games in which the computer opponent has been programmed to make rational decisions, with the goal to attain the highest possible individual pay-off, and participants are told about this.  In the games that participants play in the current paper, in contrast, the opponent often plays irrationally, and participants are told that in fact, the computer opponent is playing its best response to a strategy that it attributes to the participant. Let us see whether this leads to a difference in the application of theory of mind.  For our previous experiment presented in \cite{ghv14}, we found that higher levels of theory of mind expressed in final answers was correlated positively to number of marbles won in total in the games~\cite{Ghosh2017}, so it appears that theory of mind is indeed profitable in this type of Marble Drop games in which the opponent sometimes acts in an unexpected way by not ending the game. 

In their model-based analysis of strategies used by the participants of~\cite{meijering2011,flobbeverb}, Meijering and colleagues~\cite{meijering2014a} implemented one simple ``zero-order theory of mind" strategy  that ignores any future decisions and simply compares the immediate payoff, when stopping a game, against the maximum of all future possible payoffs, which they dubbed as a ``simple risk-taking strategy". A more advanced ``first-order theory of mind" version of that model attributes the simple risk-taking strategy to the opponent. They showed that the sequences of decisions of many 9-year old children playing the games of~\cite{flobbeverb} correspond to either the zero -order risk-taking strategy or the first-order version attributing risk-taking to the opponent; and that the adults' decisions in these games~\cite{meijering2011} correspond more closely to the first-order strategy of assigning the risk-taking strategy to the opponent, or to a second-order strategy. 

It is interesting to see in Figure~\ref{fig:finalanswer} that in our games, many participants' answers also exemplify that they themselves ``take a risk" or ``take a chance" when moving to the right, namely 40 out of 50 participants; and a similar number of participants, namely 39 out of 50, attributes such risk-taking to their opponent. Also, in line with the work by Meijering and colleagues, most of our current adult participants (34 out of 50) appear to reason explicitly at the first or second order of theory of mind, at least when their answers are analyzed. Thus, it does appear that participants regularly do think and act at least at first- or second-order theory of mind, which is rather high when considering how much more complicated our decision trees for games 1, 2, 3 and 4 are than those of~\cite{meijering2011,hedden2002} and especially~\cite{Evans2011}.

For the current experiment, it would be useful to estimate more precisely which level of theory of mind each participant actually applies when making their own decisions, using computational techniques such as simulation of computational cognitive models~\cite{ghosh2014,Ghosh2017} or Bayesian strategy estimation~\cite{Weerdpress}. 

\subsection{Moving toward cooperation}

It has long been noted in the literature on behavioral game theory that players' utilities do not correspond one-to-one to payoffs~\cite{mckelvey1992}. Healy~\cite{healy16} has elicited players' utilities over outcomes associated with centipede-like games, showing that there are many motivations that play a role. In our games, trying to get all the way to the right also corresponds to going for the outcomes with maximum social welfare, see Figures~\ref{fig:maingames} and~\ref{fig:auxgames}. It is very well possible that participants try to get there and aim to entice their opponent to move right as well (see also Nagel and Tang's experiment~\cite{nagel98}, in which participants may have reason to believe that their opponent could be an altruist).  In this light, our participants could interpret the opponent's first ``irrational" move to the right not as ``irrational" at all, but as a first step towards cooperation to attain maximum social welfare.


\section{Conclusion}\label{sec:concl}

We made a number of improvements that we thought would make it easier for participants to apply backward or forward induction reasoning in the perfect-information games in a Marble Drop set-up than in~\cite{ghv14}. We found that even though in the aggregate, participants in the new experiment still tend to slightly favor the forward induction choice at their first decision node, their verbalized strategies most often depend on their own attitudes towards risk and those they assign to the computer opponent, sometimes in addition to considerations about cooperativeness and competitiveness.
 
When we analyzed the individual participants' decisions and their answers about the reasons behind their decisions, it turns out that many players do not find anything to rationalize when the computer does not take the safe and rational option to go down and stop the game at its first decision point. Instead, in almost all the games, many participants think that the computer is just taking a chance to gain a higher payoff later. Also, many of the players are willing to take such a risk at their own first decision point: very few players take the BI option of stopping the game there. Dividing the players into classes according to their decisions as well as according to how they reasoned about their own and the opponent's choices turned up several more nuanced patterns of reasoning. 

In view of our findings for this experiment, our subsequent categorization of the elements of reasoning behind the behavior of participants, namely (i) risk-taking/risk-averseness tendency of self/opponent, (ii) competitive/cooperative tendency, and (iii) proficiency in applying higher-order theory of mind, has provided an understanding of a few possible player-types with respect to their strategic reasoning. In addition, the distinction of instinctive versus contemplative reasoners could be addressed in future investigations, requiring a detailed analysis of the time data~\cite{rubinstein2013,rubinstein2014}. Such an analysis of the temporal data could also provide new insights into the categorisation of the participants' behavior as mentioned above.

The take-home message from this paper that we would like to suggest is the following: In addition to investigating into game-theoretic rationality as a guiding force for understanding people's choices in dynamic games (both perfect and imperfect information games), one should also look into risk-taking and risk-averse behaviours, theory of mind considerations, competitive and cooperative considerations, instinctive and contemplative behaviours, and similar reasoning tendencies towards providing a better explanation of people's choices in turn-taking games.

\subsection{Acknowledgments}
We would like to thank Burkhard Schipper for his many insightful comments about our paper~\cite{ghv14} and his useful suggestions about improved ways to set up this experiment on forward versus backward induction, including ways to incentivize participants to take the games seriously. We would also like to thank Fokie Cnossen for her advice on how to formulate questions to participants.

We are vey grateful to Eric Jansen who designed and implemented the new experiment including the Marble Drop interface, which was inspired by the one constructed by Damian Podareanu and Michiel van de
Steeg for our previous experiment reported in~\cite{ghv14}. We would also like to thank Eric Jansen for performing the experiment for this study. The anonymous reviewers for TARK 2017 have given us very helpful advice to improve our paper, for which we would like to thank them.


\bibliographystyle{eptcs}
\bibliography{tark}

\begin{thebibliography}{10}
\providecommand{\bibitemdeclare}[2]{}
\providecommand{\surnamestart}{}
\providecommand{\surnameend}{}
\providecommand{\urlprefix}{Available at }
\providecommand{\url}[1]{\texttt{#1}}
\providecommand{\href}[2]{\texttt{#2}}
\providecommand{\urlalt}[2]{\href{#1}{#2}}
\providecommand{\doi}[1]{doi:\urlalt{http://dx.doi.org/#1}{#1}}
\providecommand{\bibinfo}[2]{#2}

\bibitemdeclare{article}{allais53}
\bibitem{allais53}
\bibinfo{author}{M.~\surnamestart Allais\surnameend} (\bibinfo{year}{1953}):
  \emph{\bibinfo{title}{Le comportement de l'homme rationnel devant le risque:
  {C}ritique des postulats et axiomes de l'\'{e}cole Am\'{e}ricaine}}.
\newblock {\sl \bibinfo{journal}{Econometrica}} \bibinfo{volume}{21}, pp.
  \bibinfo{pages}{503--546}, \doi{10.2307/1907921}.

\bibitemdeclare{unpublished}{healy12}
\bibitem{healy12}
\bibinfo{author}{Y.~\surnamestart Azrieli\surnameend}, \bibinfo{author}{C.~P.
  \surnamestart Chambers\surnameend} \& \bibinfo{author}{P.~J. \surnamestart
  Healy\surnameend} (\bibinfo{year}{2012}): \emph{\bibinfo{title}{Incentives in
  experiments: {A} theoretical analysis}}.
\newblock \bibinfo{note}{Working paper}.

\bibitemdeclare{article}{Bacharachstahl2000leveln}
\bibitem{Bacharachstahl2000leveln}
\bibinfo{author}{M.~\surnamestart Bacharach\surnameend} \&
  \bibinfo{author}{D.~O. \surnamestart Stahl\surnameend}
  (\bibinfo{year}{2000}): \emph{\bibinfo{title}{Variable-frame level-\emph{n}
  theory}}.
\newblock {\sl \bibinfo{journal}{Games and Economic Behavior}}
  \bibinfo{volume}{32}(\bibinfo{number}{2}), pp. \bibinfo{pages}{220--246},
  \doi{10.1006/game.2000.0796}.

\bibitemdeclare{article}{bn08}
\bibitem{bn08}
\bibinfo{author}{D.~\surnamestart Balkenborg\surnameend} \&
  \bibinfo{author}{R.~\surnamestart Nagel\surnameend} (\bibinfo{year}{2016}):
  \emph{\bibinfo{title}{An Experiment on Forward vs. Backward Induction: How
  Fairness and Level k Reasoning Matter}}.
\newblock {\sl \bibinfo{journal}{German Economic Review}}
  \bibinfo{volume}{17}(\bibinfo{number}{3}), pp. \bibinfo{pages}{378--408},
  \doi{10.1111/geer.12099}.

\bibitemdeclare{article}{battigalli96}
\bibitem{battigalli96}
\bibinfo{author}{P.~\surnamestart Battigalli\surnameend}
  (\bibinfo{year}{1996}): \emph{\bibinfo{title}{Strategic rationality orderings
  and the best rationalizability principle}}.
\newblock {\sl \bibinfo{journal}{Games and Economic Behavior}}
  \bibinfo{volume}{13}, pp. \bibinfo{pages}{178--200},
  \doi{10.1006/game.1996.0033}.

\bibitemdeclare{article}{battigalli97}
\bibitem{battigalli97}
\bibinfo{author}{P.~\surnamestart Battigalli\surnameend}
  (\bibinfo{year}{1997}): \emph{\bibinfo{title}{On rationalizability in
  extensive games}}.
\newblock {\sl \bibinfo{journal}{Journal of Economic Theory}}
  \bibinfo{volume}{74}, pp. \bibinfo{pages}{40--61},
  \doi{10.1006/jeth.1996.2252}.

\bibitemdeclare{inproceedings}{Bergwerff2014}
\bibitem{Bergwerff2014}
\bibinfo{author}{G.~\surnamestart Bergwerff\surnameend},
  \bibinfo{author}{B.~\surnamestart Meijering\surnameend},
  \bibinfo{author}{J.~\surnamestart Szymanik\surnameend},
  \bibinfo{author}{R.~\surnamestart Verbrugge\surnameend} \&
  \bibinfo{author}{S.~\surnamestart Wierda\surnameend} (\bibinfo{year}{2014}):
  \emph{\bibinfo{title}{Computational and algorithmic models of strategies in
  turn-based games}}.
\newblock In: {\sl \bibinfo{booktitle}{Proceedings of the 36th Annual
  Conference of the Cognitive Science Society}}, pp.
  \bibinfo{pages}{1778--1783}.

\bibitemdeclare{article}{cachon1996}
\bibitem{cachon1996}
\bibinfo{author}{G.~P. \surnamestart Cachon\surnameend} \&
  \bibinfo{author}{C.~\surnamestart Camerer\surnameend} (\bibinfo{year}{1996}):
  \emph{\bibinfo{title}{Loss-avoidance and forward induction in experimental
  coordination games}}.
\newblock {\sl \bibinfo{journal}{The Quarterly Journal of Economics}}
  \bibinfo{volume}{111}(\bibinfo{number}{1}), pp. \bibinfo{pages}{165--94},
  \doi{10.2307/2946661}.

\bibitemdeclare{article}{camerer2004cognitive}
\bibitem{camerer2004cognitive}
\bibinfo{author}{C.~F. \surnamestart Camerer\surnameend},
  \bibinfo{author}{T.-H. \surnamestart Ho\surnameend} \& \bibinfo{author}{J.-K.
  \surnamestart Chong\surnameend} (\bibinfo{year}{2004}):
  \emph{\bibinfo{title}{A cognitive hierarchy model of games}}.
\newblock {\sl \bibinfo{journal}{The Quarterly Journal of Economics}}
  \bibinfo{volume}{119}(\bibinfo{number}{3}), pp. \bibinfo{pages}{861--898},
  \doi{10.1162/0033553041502225}.

\bibitemdeclare{unpublished}{chenmicali11}
\bibitem{chenmicali11}
\bibinfo{author}{J.~\surnamestart Chen\surnameend} \&
  \bibinfo{author}{S.~\surnamestart Micali\surnameend} (\bibinfo{year}{2011}):
  \emph{\bibinfo{title}{The robustness of extensive-form rationalizability}}.
\newblock \bibinfo{note}{Working paper}.

\bibitemdeclare{article}{chenmicali13}
\bibitem{chenmicali13}
\bibinfo{author}{J.~\surnamestart Chen\surnameend} \&
  \bibinfo{author}{S.~\surnamestart Micali\surnameend} (\bibinfo{year}{2013}):
  \emph{\bibinfo{title}{The order independence of iterated dominance in
  extensive games}}.
\newblock {\sl \bibinfo{journal}{Theoretical Economics}} \bibinfo{volume}{8},
  pp. \bibinfo{pages}{125--163}, \doi{10.3982/TE942}.

\bibitemdeclare{article}{chlass2016}
\bibitem{chlass2016}
\bibinfo{author}{N.~\surnamestart Chla{\ss}\surnameend} \&
  \bibinfo{author}{A.~\surnamestart Perea\surnameend} (\bibinfo{year}{2016}):
  \emph{\bibinfo{title}{How do people reason in dynamic games?}}
\newblock {\sl \bibinfo{journal}{Beitr\"{a}ge zur Jahrestagung des Vereins
  f\"{u}r Socialpolitik 2016: Demographischer Wandel - Session: Economic Theory
  and Personality, No. A04-V2}}.

\bibitemdeclare{article}{costa2006cognition}
\bibitem{costa2006cognition}
\bibinfo{author}{M.~A. \surnamestart Costa-Gomes\surnameend} \&
  \bibinfo{author}{V.~P. \surnamestart Crawford\surnameend}
  (\bibinfo{year}{2006}): \emph{\bibinfo{title}{Cognition and behavior in
  two-person guessing games: An experimental study}}.
\newblock {\sl \bibinfo{journal}{The American Economic Review}}
  \bibinfo{volume}{96}(\bibinfo{number}{5}), pp. \bibinfo{pages}{1737--1768},
  \doi{10.1257/aer.96.5.1737}.

\bibitemdeclare{article}{costa2001cognition}
\bibitem{costa2001cognition}
\bibinfo{author}{M.~A. \surnamestart Costa-Gomes\surnameend},
  \bibinfo{author}{V.~P. \surnamestart Crawford\surnameend} \&
  \bibinfo{author}{B.~\surnamestart Broseta\surnameend} (\bibinfo{year}{2001}):
  \emph{\bibinfo{title}{Cognition and behavior in normal-form games: An
  experimental study}}.
\newblock {\sl \bibinfo{journal}{Econometrica}}
  \bibinfo{volume}{69}(\bibinfo{number}{5}), pp. \bibinfo{pages}{1193--1235},
  \doi{10.1111/1468-0262.00239}.

\bibitemdeclare{article}{Crawford2013}
\bibitem{Crawford2013}
\bibinfo{author}{V.~P. \surnamestart Crawford\surnameend},
  \bibinfo{author}{M.~A. \surnamestart Costa-Gomes\surnameend} \&
  \bibinfo{author}{N.~\surnamestart Iriberri\surnameend}
  (\bibinfo{year}{2013}): \emph{\bibinfo{title}{Structural models of
  nonequilibrium strategic thinking: Theory, evidence, and applications}}.
\newblock {\sl \bibinfo{journal}{Journal of Economic Literature}}
  \bibinfo{volume}{51}(\bibinfo{number}{1}), pp. \bibinfo{pages}{5--62},
  \doi{10.1257/jel.51.1.5}.

\bibitemdeclare{article}{Evans2011}
\bibitem{Evans2011}
\bibinfo{author}{A.~M. \surnamestart Evans\surnameend} \&
  \bibinfo{author}{J.~I. \surnamestart Krueger\surnameend}
  (\bibinfo{year}{2011}): \emph{\bibinfo{title}{Elements of trust: Risk and
  perspective-taking}}.
\newblock {\sl \bibinfo{journal}{Journal of Experimental Social Psychology}}
  \bibinfo{volume}{47}(\bibinfo{number}{1}), pp. \bibinfo{pages}{171--177},
  \doi{10.1016/j.jesp.2010.08.007}.

\bibitemdeclare{article}{Evans2014}
\bibitem{Evans2014}
\bibinfo{author}{A.~M. \surnamestart Evans\surnameend} \&
  \bibinfo{author}{J.~I. \surnamestart Krueger\surnameend}
  (\bibinfo{year}{2014}): \emph{\bibinfo{title}{Outcomes and expectations in
  dilemmas of trust}}.
\newblock {\sl \bibinfo{journal}{Judgment and Decision making}}
  \bibinfo{volume}{9}(\bibinfo{number}{2}), p.~\bibinfo{pages}{90}.

\bibitemdeclare{article}{flobbeverb}
\bibitem{flobbeverb}
\bibinfo{author}{L.~\surnamestart Flobbe\surnameend},
  \bibinfo{author}{R.~\surnamestart Verbrugge\surnameend},
  \bibinfo{author}{P.~\surnamestart Hendriks\surnameend} \&
  \bibinfo{author}{I.~\surnamestart Kr{\"{a}}mer\surnameend}
  (\bibinfo{year}{2008}): \emph{\bibinfo{title}{Children's Application of
  Theory of Mind in Reasoning and Language}}.
\newblock {\sl \bibinfo{journal}{Journal of Logic, Language and Information}},
  \doi{10.1007/s10849-008-9064-7}.
\newblock \bibinfo{note}{Special issue on formal models for real people, edited
  by M. Counihan}.

\bibitemdeclare{inproceedings}{ghv14}
\bibitem{ghv14}
\bibinfo{author}{S.~\surnamestart Ghosh\surnameend},
  \bibinfo{author}{A.~\surnamestart Heifetz\surnameend} \&
  \bibinfo{author}{R.~\surnamestart Verbrugge\surnameend}
  (\bibinfo{year}{2015}): \emph{\bibinfo{title}{Do players reason by forward
  induction in dynamic perfect information games?}}
\newblock In \bibinfo{editor}{R.~\surnamestart Ramanujam\surnameend}, editor:
  {\sl \bibinfo{booktitle}{Proceedings Fifteenth Conference on Theoretical
  Aspects of Rationality and Knowledge, {TARK} 2015, Carnegie Mellon
  University, Pittsburgh, USA, June 4-6, 2015.}}, {\sl
  \bibinfo{series}{{EPTCS}}} \bibinfo{volume}{215}, pp.
  \bibinfo{pages}{159--175}, \doi{10.4204/EPTCS.215.12}.
\newblock \urlprefix\url{https://doi.org/10.4204/EPTCS.215.12}.

\bibitemdeclare{article}{ghosh2014}
\bibitem{ghosh2014}
\bibinfo{author}{S.~\surnamestart Ghosh\surnameend},
  \bibinfo{author}{B.~\surnamestart Meijering\surnameend} \&
  \bibinfo{author}{R.~\surnamestart Verbrugge\surnameend}
  (\bibinfo{year}{2014}): \emph{\bibinfo{title}{Strategic Reasoning: Building
  Cognitive Models from Logical Formulas}}.
\newblock {\sl \bibinfo{journal}{Journal of Logic, Language and Information}}
  \bibinfo{volume}{23}(\bibinfo{number}{1}), pp. \bibinfo{pages}{1--29},
  \doi{10.1007/s10849-014-9196-x}.

\bibitemdeclare{article}{Ghosh2017}
\bibitem{Ghosh2017}
\bibinfo{author}{S.~\surnamestart Ghosh\surnameend} \&
  \bibinfo{author}{R.~\surnamestart Verbrugge\surnameend}
  (\bibinfo{year}{2017}): \emph{\bibinfo{title}{Studying strategies and types
  of players: Experiments, logics and cognitive models}}.
\newblock {\sl \bibinfo{journal}{Synthese}}, pp. \bibinfo{pages}{1--43},
  \doi{10.1007/s11229-017-1338-7}.

\bibitemdeclare{article}{Gill2016}
\bibitem{Gill2016}
\bibinfo{author}{D.~\surnamestart Gill\surnameend} \&
  \bibinfo{author}{V.~\surnamestart Prowse\surnameend} (\bibinfo{year}{2016}):
  \emph{\bibinfo{title}{Cognitive ability, character skills, and learning to
  play equilibrium: A level-k analysis}}.
\newblock {\sl \bibinfo{journal}{Journal of Political Economy}}
  \bibinfo{volume}{124}(\bibinfo{number}{6}), pp. \bibinfo{pages}{1619--1676},
  \doi{10.1086/688849}.

\bibitemdeclare{article}{goeree2004model}
\bibitem{goeree2004model}
\bibinfo{author}{J.~K. \surnamestart Goeree\surnameend} \&
  \bibinfo{author}{Ch.~A. \surnamestart Holt\surnameend}
  (\bibinfo{year}{2004}): \emph{\bibinfo{title}{A model of noisy
  introspection}}.
\newblock {\sl \bibinfo{journal}{Games and Economic Behavior}}
  \bibinfo{volume}{46}(\bibinfo{number}{2}), pp. \bibinfo{pages}{365--382},
  \doi{10.1016/S0899-8256(03)00145-3}.

\bibitemdeclare{unpublished}{healy16}
\bibitem{healy16}
\bibinfo{author}{P.~J. \surnamestart Healy\surnameend} (\bibinfo{year}{2016}):
  \emph{\bibinfo{title}{Epistemic experiments: Utilities, beliefs, and
  irrational play}}.
\newblock
  \urlprefix\url{http://healy.econ.ohio-state.edu/present/Healy-EpistemicPresentation90min.pdf}.
\newblock \bibinfo{note}{Unpublished presentation}.

\bibitemdeclare{article}{hedden2002}
\bibitem{hedden2002}
\bibinfo{author}{T.~\surnamestart Hedden\surnameend} \&
  \bibinfo{author}{J.~\surnamestart Zhang\surnameend} (\bibinfo{year}{2002}):
  \emph{\bibinfo{title}{What do you think {I} think you think?: Strategic
  reasoning in matrix games}}.
\newblock {\sl \bibinfo{journal}{Cognition}}
  \bibinfo{volume}{85}(\bibinfo{number}{1}), pp. \bibinfo{pages}{1--36},
  \doi{10.1016/S0010-0277(02)00054-9}.

\bibitemdeclare{article}{hp14}
\bibitem{hp14}
\bibinfo{author}{A.~\surnamestart Heifetz\surnameend} \&
  \bibinfo{author}{A.~\surnamestart Perea\surnameend} (\bibinfo{year}{2015}):
  \emph{\bibinfo{title}{On the outcome equivalence of backward induction and
  extensive form rationalizability}}.
\newblock {\sl \bibinfo{journal}{International Journal of Game Theory}}
  \bibinfo{volume}{44}(\bibinfo{number}{1}), pp. \bibinfo{pages}{37--59},
  \doi{10.1007/s00182-014-0418-x}.

\bibitemdeclare{article}{huck2005}
\bibitem{huck2005}
\bibinfo{author}{S.~\surnamestart Huck\surnameend} \&
  \bibinfo{author}{W.~\surnamestart M{\"u}ller\surnameend}
  (\bibinfo{year}{2005}): \emph{\bibinfo{title}{Burning money and (pseudo)
  first-mover advantages: An experimental study on forward induction}}.
\newblock {\sl \bibinfo{journal}{Games and Economic Behavior}}
  \bibinfo{volume}{51}(\bibinfo{number}{1}), pp. \bibinfo{pages}{109--127},
  \doi{10.1016/j.geb.2004.03.006}.

\bibitemdeclare{article}{Kawagoe2012}
\bibitem{Kawagoe2012}
\bibinfo{author}{T.~\surnamestart Kawagoe\surnameend} \&
  \bibinfo{author}{H.~\surnamestart Takizawa\surnameend}
  (\bibinfo{year}{2012}): \emph{\bibinfo{title}{Level-k analysis of
  experimental centipede games}}.
\newblock {\sl \bibinfo{journal}{Journal of Economic Behavior and
  Organization}} \bibinfo{volume}{82}(\bibinfo{number}{2}), pp.
  \bibinfo{pages}{548--566}, \doi{10.1016/j.jebo.2012.03.010}.

\bibitemdeclare{book}{mccutcheon1987latent}
\bibitem{mccutcheon1987latent}
\bibinfo{author}{A.~L. \surnamestart McCutcheon\surnameend}
  (\bibinfo{year}{1987}): \emph{\bibinfo{title}{Latent Class Analysis}}.
\newblock {\sl \bibinfo{series}{Sage University Paper Series on Quantitative
  Applications in the Social Sciences}} \bibinfo{volume}{07-064},
  \bibinfo{publisher}{Sage Publications}, \bibinfo{address}{Newbury Park, CA},
  \doi{10.4135/9781412984713}.

\bibitemdeclare{article}{mckelvey1992}
\bibitem{mckelvey1992}
\bibinfo{author}{R.~D. \surnamestart McKelvey\surnameend} \&
  \bibinfo{author}{T.~R. \surnamestart Palfrey\surnameend}
  (\bibinfo{year}{1992}): \emph{\bibinfo{title}{An experimental study of the
  centipede game}}.
\newblock {\sl \bibinfo{journal}{Econometrica: Journal of the Econometric
  Society}}, pp. \bibinfo{pages}{803--836}, \doi{10.2307/2951567}.

\bibitemdeclare{article}{mckelvey1995quantal}
\bibitem{mckelvey1995quantal}
\bibinfo{author}{R.~D. \surnamestart McKelvey\surnameend} \&
  \bibinfo{author}{T.~R. \surnamestart Palfrey\surnameend}
  (\bibinfo{year}{1995}): \emph{\bibinfo{title}{Quantal response equilibria for
  normal form games}}.
\newblock {\sl \bibinfo{journal}{Games and Economic Behavior}}
  \bibinfo{volume}{10}(\bibinfo{number}{1}), pp. \bibinfo{pages}{6--38},
  \doi{10.1006/game.1995.1023}.

\bibitemdeclare{inproceedings}{meijering2010}
\bibitem{meijering2010}
\bibinfo{author}{B.~\surnamestart Meijering\surnameend},
  \bibinfo{author}{L.~\surnamestart van Maanen\surnameend},
  \bibinfo{author}{H.~\surnamestart van Rijn\surnameend} \&
  \bibinfo{author}{R.~\surnamestart Verbrugge\surnameend}
  (\bibinfo{year}{2010}): \emph{\bibinfo{title}{The facilitative effect of
  context on second-order social reasoning}}.
\newblock In: {\sl \bibinfo{booktitle}{Proceedings of the 32nd Annual
  Conference of the Cognitive Science Society}}, pp.
  \bibinfo{pages}{1423--1428}.

\bibitemdeclare{inproceedings}{meijering2011}
\bibitem{meijering2011}
\bibinfo{author}{B.~\surnamestart Meijering\surnameend},
  \bibinfo{author}{H.~\surnamestart van Rijn\surnameend},
  \bibinfo{author}{N.~A. \surnamestart Taatgen\surnameend} \&
  \bibinfo{author}{R.~\surnamestart Verbrugge\surnameend}
  (\bibinfo{year}{2011}): \emph{\bibinfo{title}{I do know what you think {I}
  think: Second-order theory of mind in strategic games is not that
  difficult}}.
\newblock In: {\sl \bibinfo{booktitle}{Proc. 33rd Annual Conf. Cognitive
  Science Society}}, pp. \bibinfo{pages}{2486--2491}.

\bibitemdeclare{article}{meijering2012}
\bibitem{meijering2012}
\bibinfo{author}{B.~\surnamestart Meijering\surnameend},
  \bibinfo{author}{H.~\surnamestart van Rijn\surnameend}, \bibinfo{author}{N.A.
  \surnamestart Taatgen\surnameend} \& \bibinfo{author}{R.~\surnamestart
  Verbrugge\surnameend} (\bibinfo{year}{2012}): \emph{\bibinfo{title}{What eye
  movements can tell about theory of mind in a strategic game}}.
\newblock {\sl \bibinfo{journal}{PloS ONE}}
  \bibinfo{volume}{7}(\bibinfo{number}{9}), p. \bibinfo{pages}{e45961},
  \doi{10.1371/journal.pone.0045961}.

\bibitemdeclare{article}{meijering2014a}
\bibitem{meijering2014a}
\bibinfo{author}{B.~\surnamestart Meijering\surnameend}, \bibinfo{author}{N.~A.
  \surnamestart Taatgen\surnameend}, \bibinfo{author}{H.~\surnamestart van
  Rijn\surnameend} \& \bibinfo{author}{R.~\surnamestart Verbrugge\surnameend}
  (\bibinfo{year}{2014}): \emph{\bibinfo{title}{Modeling inference of mental
  states: As simple as possible, as complex as necessary}}.
\newblock {\sl \bibinfo{journal}{Interaction Studies}}
  \bibinfo{volume}{15}(\bibinfo{number}{3}), pp. \bibinfo{pages}{455--477},
  \doi{10.1075/is.15.3.05mei}.

\bibitemdeclare{article}{Nagel1995}
\bibitem{Nagel1995}
\bibinfo{author}{R.~\surnamestart Nagel\surnameend} (\bibinfo{year}{1995}):
  \emph{\bibinfo{title}{Unraveling in guessing games: An experimental study}}.
\newblock {\sl \bibinfo{journal}{The American Economic Review}}
  \bibinfo{volume}{85}(\bibinfo{number}{5}), pp. \bibinfo{pages}{1313--1326}.

\bibitemdeclare{article}{nagel98}
\bibitem{nagel98}
\bibinfo{author}{R.~\surnamestart Nagel\surnameend} \& \bibinfo{author}{F.~F.
  \surnamestart Tang\surnameend} (\bibinfo{year}{1998}):
  \emph{\bibinfo{title}{Experimental results on the centipede game in normal
  form: An investigation on learning}}.
\newblock {\sl \bibinfo{journal}{Journal of Mathematical Psychology}}
  \bibinfo{volume}{42}(\bibinfo{number}{2}), pp. \bibinfo{pages}{356--384},
  \doi{10.1006/jmps.1998.1225}.

\bibitemdeclare{incollection}{Pacuit2015}
\bibitem{Pacuit2015}
\bibinfo{author}{E.~\surnamestart Pacuit\surnameend} (\bibinfo{year}{2015}):
  \emph{\bibinfo{title}{Dynamic models of rational deliberation in games}}.
\newblock In \bibinfo{editor}{J.~van \surnamestart Benthem\surnameend},
  \bibinfo{editor}{S.~\surnamestart Ghosh\surnameend} \&
  \bibinfo{editor}{R.~\surnamestart Verbrugge\surnameend}, editors: {\sl
  \bibinfo{booktitle}{Models of Strategic Reasoning: Logics, Games and
  Communities}}, {\sl \bibinfo{series}{LNCS-FoLLI}} \bibinfo{volume}{8972},
  \bibinfo{publisher}{Springer}, \bibinfo{address}{New York, NY}, pp.
  \bibinfo{pages}{3--33}, \doi{10.1007/978-3-662-48540-8\textunderscore1}.

\bibitemdeclare{article}{pearce84}
\bibitem{pearce84}
\bibinfo{author}{D.~\surnamestart Pearce\surnameend} (\bibinfo{year}{1984}):
  \emph{\bibinfo{title}{Rationalizable strategic behaviour and the problem of
  perfection}}.
\newblock {\sl \bibinfo{journal}{Econometrica}} \bibinfo{volume}{52}, pp.
  \bibinfo{pages}{1029--1050}, \doi{10.2307/1911197}.

\bibitemdeclare{incollection}{perea07}
\bibitem{perea07}
\bibinfo{author}{A.~\surnamestart Perea\surnameend} (\bibinfo{year}{2007}):
  \emph{\bibinfo{title}{Epistemic foundations for backward induction: An
  overview}}.
\newblock In \bibinfo{editor}{J.~\surnamestart van Benthem\surnameend},
  \bibinfo{editor}{D.~\surnamestart Gabbay\surnameend} \&
  \bibinfo{editor}{B.~\surnamestart L{\"o}we\surnameend}, editors: {\sl
  \bibinfo{booktitle}{Interactive Logic}}, {\sl \bibinfo{series}{Texts in Logic
  and Games}}~\bibinfo{volume}{1}, \bibinfo{publisher}{Amsterdam University
  Press}, pp. \bibinfo{pages}{159--193}.

\bibitemdeclare{book}{perea12}
\bibitem{perea12}
\bibinfo{author}{A.~\surnamestart Perea\surnameend} (\bibinfo{year}{2012}):
  \emph{\bibinfo{title}{Epistemic Game Theory: Reasoning and Choice}}.
\newblock \bibinfo{publisher}{Cambridge University Press},
  \bibinfo{address}{Cambridge}, \doi{10.1017/CBO9780511844072}.

\bibitemdeclare{article}{Premack1978}
\bibitem{Premack1978}
\bibinfo{author}{D.~\surnamestart Premack\surnameend} \&
  \bibinfo{author}{G.~\surnamestart Woodruff\surnameend}
  (\bibinfo{year}{1978}): \emph{\bibinfo{title}{Does the chimpanzee have a
  theory of mind?}}
\newblock {\sl \bibinfo{journal}{Behavioral and Brain Sciences}}
  \bibinfo{volume}{1}(\bibinfo{number}{04}), pp. \bibinfo{pages}{515--526},
  \doi{10.1017/S0140525X00076512}.

\bibitemdeclare{article}{reny92}
\bibitem{reny92}
\bibinfo{author}{P.~J. \surnamestart Reny\surnameend} (\bibinfo{year}{1992}):
  \emph{\bibinfo{title}{Backward induction, normal form perfection and
  explicable equilibria}}.
\newblock {\sl \bibinfo{journal}{Econometrica}} \bibinfo{volume}{60}, pp.
  \bibinfo{pages}{627--649}, \doi{10.2307/2951586}.

\bibitemdeclare{article}{rubinstein2013}
\bibitem{rubinstein2013}
\bibinfo{author}{A.~\surnamestart Rubinstein\surnameend}
  (\bibinfo{year}{2013}): \emph{\bibinfo{title}{Response time and decision
  making: An experimental study}}.
\newblock {\sl \bibinfo{journal}{Judgment and Decision Making}}
  \bibinfo{volume}{8}(\bibinfo{number}{5}), pp. \bibinfo{pages}{540--551},
  \doi{10.1.1.396.668}.

\bibitemdeclare{article}{rubinstein2014}
\bibitem{rubinstein2014}
\bibinfo{author}{A.~\surnamestart Rubinstein\surnameend}
  (\bibinfo{year}{2016}): \emph{\bibinfo{title}{A typology of players: Between
  instinctive and contemplative}}.
\newblock {\sl \bibinfo{journal}{Quarterly Journal of Economics}}
  \bibinfo{volume}{131}(\bibinfo{number}{2}), pp. \bibinfo{pages}{859--890},
  \doi{10.1093/qje/qjw008}.

\bibitemdeclare{article}{shahriar14}
\bibitem{shahriar14}
\bibinfo{author}{Q.~\surnamestart Shahriar\surnameend} (\bibinfo{year}{2014}):
  \emph{\bibinfo{title}{An experimental test of the robustness and the power of
  forward induction}}.
\newblock {\sl \bibinfo{journal}{Managerial and Decision Economics}}
  \bibinfo{volume}{35}, pp. \bibinfo{pages}{264--277}, \doi{10.1002/mde.2613}.

\bibitemdeclare{article}{stahl1995players}
\bibitem{stahl1995players}
\bibinfo{author}{D.~O. \surnamestart Stahl\surnameend} \&
  \bibinfo{author}{P.~W. \surnamestart Wilson\surnameend}
  (\bibinfo{year}{1995}): \emph{\bibinfo{title}{On players' models of other
  players: Theory and experimental evidence}}.
\newblock {\sl \bibinfo{journal}{Games and Economic Behavior}}
  \bibinfo{volume}{10}(\bibinfo{number}{1}), pp. \bibinfo{pages}{218--254},
  \doi{10.1006/game.1995.1031}.

\bibitemdeclare{article}{Weerd2017}
\bibitem{Weerd2017}
\bibinfo{author}{H.~\surnamestart de~Weerd\surnameend},
  \bibinfo{author}{R.~\surnamestart Verbrugge\surnameend} \&
  \bibinfo{author}{B.~\surnamestart Verheij\surnameend} (\bibinfo{year}{2017}):
  \emph{\bibinfo{title}{Negotiating with other minds: The role of recursive
  theory of mind in negotiation with incomplete information}}.
\newblock {\sl \bibinfo{journal}{Autonomous Agents and Multi-Agent Systems}}
  \bibinfo{volume}{31}(\bibinfo{number}{2}), pp. \bibinfo{pages}{250--287},
  \doi{10.1007/s10458-015-9317-1}.

\bibitemdeclare{article}{Weerdpress}
\bibitem{Weerdpress}
\bibinfo{author}{H.~de \surnamestart Weerd\surnameend},
  \bibinfo{author}{D.~\surnamestart Diepgrond\surnameend} \&
  \bibinfo{author}{R.~\surnamestart Verbrugge\surnameend} (\bibinfo{year}{2017,
  to appear}): \emph{\bibinfo{title}{Estimating the use of higher-order theory
  of mind using computational agents}}.
\newblock {\sl \bibinfo{journal}{The B.E. Journal of Theoretical Economics}}.

\bibitemdeclare{article}{Weerd2013}
\bibitem{Weerd2013}
\bibinfo{author}{H.~de \surnamestart Weerd\surnameend},
  \bibinfo{author}{R.~\surnamestart Verbrugge\surnameend} \&
  \bibinfo{author}{B.~\surnamestart Verheij\surnameend} (\bibinfo{year}{2013}):
  \emph{\bibinfo{title}{How much does it help to know what she knows you know?
  An agent-based simulation study}}.
\newblock {\sl \bibinfo{journal}{Artificial Intelligence}}
  \bibinfo{volume}{199}, pp. \bibinfo{pages}{67--92},
  \doi{10.1016/j.artint.2013.05.004}.

\bibitemdeclare{inproceedings}{wright2010beyond}
\bibitem{wright2010beyond}
\bibinfo{author}{J.~R. \surnamestart Wright\surnameend} \&
  \bibinfo{author}{K.~\surnamestart Leyton-Brown\surnameend}
  (\bibinfo{year}{2010}): \emph{\bibinfo{title}{Beyond equilibrium: Predicting
  human behavior in normal-form games}}.
\newblock In: {\sl \bibinfo{booktitle}{Proceedings of the 24th Conference on
  Artificial Intelligence}}, pp. \bibinfo{pages}{901--907}.

\end{thebibliography}

\newpage
\section*{Appendix A: Instruction sheet for the participants}

Here follows the written information on the instruction sheet that was given to each participant and explained by the experimenter before the start of the training games.

\begin{itemize}
\item[-] In this task, you will be playing two-player games. The computer is the other player.
\item[-] In each game, a purple marble is about to drop, and both you and the computer determine its path by controlling the orange and the blue trapdoors.
\item[-] You control the orange trapdoors, and the computer controls the blue trapdoors.
\item[-] Your goal is that the purple marble drops into the bin with as many orange marbles as possible. The computer's goal is that the marble drops into the bin with as many blue marbles as possible.
\item[-] Click on the left trapdoor if you want the marble to go left, and on the right trapdoor if you want the marble to go right.
\item[-] How the computer reasons in each particular game?
\begin{quote}
The computer thinks that you already have a plan for that game, and it plays the best response to the plan it thinks that you have for that game.

However, the computer does not learn from previous games and does not take into account your choices during the previous games.
\end{quote}
\item[-] The first 14 games are practice games. At the end of each practice game, you will see how many marbles you collected in that game.
\item[-] The practice games are followed by 48 experiment games. At the end of each experiment game, you will see how many marbles you collected in that game.
\item[-] You will be able to start each game by clicking on the ``START GAME'' button, and move to the next game by clicking on the ``NEXT'' button.
\item[-] At some points during the experiment phase, you will be asked a few questions regarding your choices. Please think carefully about each question before answering it.
\item[-] There will be a break of 5 minutes once you finish 24 of the 48 experiment games.
\item[-] After the the 48 experiment games, you will be asked a few final questions regarding 2 example games.
\item[-] The money you will earn is between \EUR{3.75} and \EUR{15} and depends on the marbles you have won during the experiment phase. At the end of the experiment, 1 of the 48 experiment games will be chosen at random. Games where you were not given at least one choice will be omitted from this random selection. The number of marbles that you have collected in the randomly selected game will determine your final earnings: each collected marble is worth \EUR{3.75}.

\end{itemize}


\end{document}